\documentclass[amstex,epsf,amssymb,usenatbib]{mnras}
\usepackage{epsf,graphics}
\usepackage{amsmath,amssymb,natbib}
\usepackage{lmodern}
\usepackage{cite}

\usepackage{rotating,times,graphicx,latexsym,color,longtable,lscape} 

\def\aj{AJ}                   
\def\apj{ApJ}                 
\def\apjl{ApJ}                
\def\apjs{ApJS}               
\def\aap{A\&A}                
\def\mnras{MNRAS}             
\def\nat{Nature}              

\newcommand{\degsy}{$^\circ$}

\title{The physical structure of  radio galaxies explored with three-dimensional simulations}
    
\author[J. Donohoe \& M.D. Smith]
{Justin Donohoe $^{1}$\thanks{E-mail: jd440@kent.ac.uk, }
\& {Michael D. Smith $^{1}$\thanks{E-mail: m.d.smith@kent.ac.uk}}\\
$^{1}$Centre for Astrophysics \& Planetary Science, The University of Kent, Canterbury, Kent CT2 7NH, U.K. }                                                                                                                                                             
\date{Accepted .....
      Received ..... ;
      in original form .....}
                                                                                                                                                             
\pagerange{\pageref{firstpage}--\pageref{lastpage}}
\pubyear{2014}
                                                                                                                                                             
\begin{document}
                                                                                                                                                             
\maketitle
                                                                                                                                                             
\label{firstpage}
                                                                                                                                                             
\begin{abstract}

\noindent We present a large systematic study of hydrodynamic simulations of supersonic adiabatic jets in three dimensions to provide a definitive set of results on exploring jet density, Mach number and precession angle as variables. We restrict the set-up to non-relativistic pressure-equilibrium flows into a homogeneous environment. We first focus on the distribution and evolution of physical parameters associated with radio galaxies. We find that the jet density has limited influence on the structure for a given jet Mach number. The speed of advance varies by a small factor for jet densities between 0.1 and 0.0001 of the ambient density while the cocoon and cavity evolution change from narrow pressure balanced to wide over-pressure as the ratio falls. We also find that the fraction of energy transferred to the ambient medium increases with decreasing jet-ambient density ratio, reaching $\approx$ 80\%. This energy is predominantly in thermal energy with almost all the remainder in ambient kinetic form. The total energy remaining in the lobe is typically under 5\%. We find that radio galaxies with wide transverse cocoons can be generated through slow  precession at low Mach numbers. 
We explore a  slow precession model in which the jet direction changes very slowly relative to the jet flow dynamical time. This reveals two separated bow shocks propagating into the ambient medium, one associated with the entire lobe expansion and the other with the immediate impact zone. The lobes generated  are generally consistent with observations, displaying  straight jets but asymmetric lobes.  
 
   \end{abstract}
  
\section{Introduction}              
\label{intro}
   
Radio galaxies are some of the largest \citep{1974Natur.250..625W,2013MNRAS.432..200M} and most powerful structures \citep{2009A&A...494..471O} that we observe in the universe. Their power is derived from processes in the vicinity of supermassive black holes which lie at the heart of active galactic nuclei \citep{1982Natur.295...17R,1991Natur.349..138R}. Their power is channelled out through jets containing an unknown  mixture of high-energy plasma, relativistic particles and  magnetic field, as deduced from  radio hot spots and lobes where a small fraction of the energy  is converted into synchrotron radiation \citep[cf.][]{2012ajb..book.....S}. They are also cosmologically important as evolving beacons at high redshift \citep{2013MNRAS.432..200M}, as regulators of galaxy growth and with  implications for processes involved with black hole physics \citep{2012ajb..book.....S}.

There are many types of radio galaxies and much effort has gone into the classification system and associated dynamical processes. Classical examples display approximate jet-axial mirror symmetry (in addition to twin-lobe mirror symmetry) which has justified the use of two-dimensional numerical simulations \citep[e.g.][]{2002ApJS..141..337C} although it is unlikely that such simulations can accurately reproduce the consequences of turbulence and fluid instabilities  \citep{2002ApJS..141..337C} even in this context \citep{2001A&A...380..789K}.  Moreover, the first  such adiabatic hydrodynamic study demonstrated that the shape of the cocoon and hot spots continuously change structure \citep{1985MNRAS.214...67S} as the pressure of the  expelled gas feeds back onto the approaching jet \citep{1982A&A...113..285N}.  Nevertheless, \citet{2013MNRAS.430..174H} demonstrated that two-dimensional calculations can provide the framework in which to discuss the overall energy and pressure distributions within a non-uniform environment corresponding to propagation through a galactic medium. The  associated high spatial resolution that can be achieved in two dimensions also permits the simulations to be run to greater jet lengths.

A systematic study in three dimensions at high resolution may help verify or contradict the trends found in two-dimensional simulations. We expect symmetry to be broken by jet precession \citep{1978Natur.276..588E} or reorientation \citep{2012ApJ...746..167H}, interstellar pressure gradients \citep{1981MNRAS.194..771S}, intracluster gas motions \citep{1979Natur.279..770B}, inter-cluster gas dynamics \citep{1995ApJ...445...80L} or simply through  instability at the impact interface. 

However, with only a few three dimensional studies performed, many properties have yet to be systematically investigated. The first attempts were directed at jet bending \citep{1992ApJ...393..631B,1995ApJ...445...80L}.  \citet{2005ApJ...633..717O} found that energy is transferred efficiently to the ambient medium with approximately half of the jet energy being converted to thermal energy in the ambient medium. \citet{2012ApJ...750..166M} analysed three-dimensional magnetohydrodynamical simulations of intermittent jets in a turbulent intracluster medium. This weather distorts the jets and lobes into morphologies similar to wide-angle tail radio sources while the intermittency leads to features similar to those in double-double radio galaxies. Finally, \citet{2014MNRAS.443.1482H} also consider a cluster environment, taking a low Mach number jet. They demonstrate the polarisation properties and show a major difference between synchrotron and inverse-Compton emission images. 

In this paper, we are  interested in the dynamical state of the injected jets and the corresponding physical structures generated rather than the influence of the ambient medium. In a subsequent paper, we investigate the observational predictions including the associated X-ray cavity and the role of the angle of orientation to the line of sight. 

In this programme, the ultimate goal is to provide a set of radio and X-ray images of radio galaxies at various orientations and distances from the observer. How radio galaxies appear at high redshift and close to the line of sight will provide a means of interpretation of data from the next generation of telescopes. Can we deduce the causes of observed features, and constrain the mass, momentum and energy budgets? To achieve this, we first require the dependence on the intrinsic flow parameters to be established before proceeding in a follow-up work to present the morphologies. 

Fundamental injection parameters which have yet to be systematically studied within three dimensional simulations include the density, precession angle and the Mach number. The first issue we confront here is the dependence on the jet-ambient density ratio, $\eta$.The density ratio is thought to be critical to the radio galaxy morphology with lower density ratios generating wider lobes in a self-similar analysis which assumes the lobes maintain a high pressure in comparison to the ambient medium. However, it is not supported by the two dimensional simulations  of \citet{2003A&A...398..113K}. In addition, the reflection boundary condition along the plane containing the jet nozzle is crucial since the cocoon material is trapped on the grid and a wide cavity builds up; whereas with an outflow condition, the cocoon remains quite cylindrical and uniform in time. \citet{2013MNRAS.430..174H}  discuss the relevance of the self-similar solutions \citep{1989ApJ...345L..21B}, as first raised by \citet{1974MNRAS.166..513S}, in which the jet-supplied cocoon maintains a high pressure and concludes that it can be dismissed on observational grounds. This is because, after a brief initial stage, the cocoon material has time to expand and readily reaches approximate pressure balance with the ambient medium. 

Secondly, the jet Mach number is a crucial parameter as first realised by \citet{1983ASSL..103..227N} and studied by \citet{1985PASAu...6..130B}. The latter work explored the possibility that the Mach number determines whether a radio source is of type FR-I or FR-II as defined by \citet{1974MNRAS.167P..31F}. 
If hot spots are present in FR-Is,  they occur close to the host galaxy. whilst FR-IIs have prominent hot spots that occur further away from the host galaxy  
FR-I radio galaxies tend to have visible jets whilst FR-II  jets tend to be faint id at all detectable.  Holding the jet density constant, \citet{2005ApJ...633..717O} looked at the influence of jet Mach number and the stratified ambient medium within five runs. They found that low Mach number jets are somewhat more easily disrupted.

Thirdly, the smooth re-alignment, wobbling or precessing of the jet direction has not been greatly explored in the context of giant radio galaxies. While crucial to the structures formed by heavy (ballistic) jets, the short jet dynamical time in comparison to the lobe dynamical time for light jets implies that precession-related phenomena may be more difficult to identify \citep{2011ApJ...734L..32G}. Thus, rather than generating a helical structure, we should observe curved or arcuate lobes as if a paint brush has swept across a canvas \citep{1978Natur.276..588E} or a distinct X-shaped structure \citep{2007AJ....133.2097C}. A combination of re-alignment and intracluster wind may be necessary \citep{2011ApJ...733...58H}.

We limit the present study to that of a non-relativistic adiabatic flow of a uniform supersonic jet. The jet is injected into a uniform stationary ambient medium. It begins perfectly collimated from a nozzle with a simulated circular cross-section. We do not include  pulsations, shear or an orbiting nozzle. In addition, magnetic and gravitational forces are also ignored.

\begin{table*}
     \caption[table1]
      {The initial conditions: both non-dimensional parameters and their example scaled interpretations taking a jet with a density of 1\% of the ambient density. The parameter $n_\text{p,amb}$,is the hydrogen nuclei(free proton) density in the ambient medium. }
      \label{parameters}
   \begin{tabular}{lrrr}
           \hline  \noalign{\smallskip}
        \sf{-}   &  \sf{unit}  & \sf{Compact} & \sf{Giant}      \\
         \sf{-}  &  \sf{ value}                   &   \sf{Source}  &  \sf{Source}    \\
              \noalign{\smallskip} \hline     \noalign{\smallskip}
         $ D$                       &   30      &     75.0~kpc        &  750~kpc     \\
          $ r_\text{jet}$         &    1      &       2.5~kpc         &   25~kpc      \\
          $M_\text{jet}$          &    6.0    &        6.0        &   6.0        \\         
          $\rho_\text{amb}$       &    1      &  2.34~10$^{-26}$ ~g~cm$^{-3}$  &   9.37~10$^{-28}$ ~g~cm$^{-3}$  \\  
          $c_\text{amb}$          &    1      &  6.72~10$^{7}$~~~~cm~s$^{-1}$     &   8.23~10$^{7}$~~~~cm~s$^{-1}$     \\          
    & & & \\       
            $n_\text{p,amb}$       &   n/a   &   1.0~10$^{-2}$ ~cm$^{-3}$      &  4.0~10$^{-4}$ ~cm$^{-3}$  \\  
            $ T_\text{amb}$        &  n/a    &     2.0~10$^7$~~~~~K~~~~~~~                  &  3.0~10$^7$~~~~~K~~~~~~~      \\         
             $u_\text{amb}$        &   0.9   &        9.53~10$^{-11}$~~~erg~cm$^{-3}$                &    5.72~10$^{-12}$ ~~~erg~cm$^{-3}$          \\    
            $p_\text{amb}$         &   0.6   &         6.35~10$^{-11}$~dyne~cm$^{-2}$                  &   3.81~10$^{-12}$~dyne~cm$^{-2}$       \\             
     & & & \\
            $ p_\text{jet}/p_\text{amb}$ &    1.00   &      1.00     &    1.00    \\     
      $ \rho_\text{jet}/\rho_\text{amb}$ &    0.01  &      0.01    &    0.01   \\              

             $v_\text{jet}$           &    60.0   &     4.03~10$^{9}$~cm~s$^{-1}$      &  4.94~10$^{9}$~cm~s$^{-1}$         \\   
              $\dot {\mathcal{M}}_\text{jet}$  &    1.88   &         2.80~M$_\odot$~yr$^{-1}$ &  13.7~M$_\odot$~yr$^{-1}$      \\            
             $P_\text{ram}$          &    113   &      7.12~10$^{35}$~dyne~~~~~          &   4.27~10$^{36}$~dyne~~~~~        \\   
             $L_\text{jet}$       &     3,562   &        1.51~10$^{45}$~erg~s$^{-1}$          &  1.11~10$^{46}$~erg~s$^{-1}$         \\   
              $t _{o}= r_\text{jet}/c_\text{amb}$            &      1         &       3.64~Myr   &           29.7~Myr \\     
                $t _\text{precession}= 2{\pi}/\omega$    &      4         &     14.64~Myr   &          118.8~Myr \\                           
             $t _\text{lobe-dynamic}= D/U$                          &      5         &       18.18~Myr   &          148.4~Myr \\    
             $t _\text{jet-dynamic}= D/v_\text{jet}$   &      0.5         &       1.82~Myr   &          14.8~Myr \\   
        \noalign{\smallskip}  \hline  \noalign{\smallskip}
   \end{tabular}
\end{table*}

\section{Method}
\label{Method}

\subsection{The Codes}
 
Two efficient codes for computational fluid dynamics, suitable for simulation surveys,  are tested and employed. ZEUS-3D is a grid-based  second-order Eulerian finite difference code \citep{1992ApJS...80..753S}  that uses Van Leer advection and consistent transport of the magnetic field. With von Neumann and Richtmyer artificial viscosity and an upwinded scheme, it is ideal for problems involving supersonic flow and is versatile, robust and well-tested \citep{2010ApJS..187..119C}. Although higher order codes are potentially more accurate, the high speed of the algorithms means that large   problems can be solved at high resolution.

We employ version 3.5 of ZEUS-3D (dzeus35) which is freely available for use by the scientific community and can be downloaded from the Institute of Computational Astrophysics (ICA) at St.Mary's University, Nova Scotia, Canada\,\footnote{\tt http://www.ica.smu.ca/zeus3d/}.  
 
 PLUTO is similarly grid-based but incorporates  modern Godunov-type shock-capturing schemes \citep{2007ApJS..170..228M}. After comparing the results of  numerous options, we chose a fast linear interpolation Hancock time-stepping (denoted HLLC) scheme\,\footnote{\tt http://plutocode.ph.unito.it/}.  

For the simulations that involve only adiabatic hydrodynamics (HD), six properties are recorded to file: the density, $\rho$, pressure, $p$, three velocity components, $v_x$, $v_y$ and $v_z$,  and a mass-weighted jet tracer, $\chi$. 

\subsection{Scaling}\label{Scaling}

In order to adequately  cover wide precessing radio sources, we base the simulations on a $D^3$ = 30x30x30 unit volume where the jet radius, $r_\text{jet}$, is set to one unit. The ambient medium is taken to be uniform with a sound speed, $c_\text{amb}$, of one unit. This sets the time scale, $r_{\text jet}/c_\text{amb}$, also to one unit.  Given an ambient density of one unit and
\begin{equation}\label{eq_camb}
        c_{\text{amb}} =  \sqrt{ \gamma . p_{\text{amb}} \over \rho_{\text{amb}}}, 
\end{equation}
yields a pressure $p_{\text amb}$ = 0.6 and internal energy per unit volume $u_{\text amb}$ = 0.9 for the specific heat ratio of $\gamma$ = 5/3 since
\begin{equation}\label{eq_pamb}
         p_{\text{amb}} =  (\gamma - 1) u_{\text{amb}}.
\end{equation}
We assume adiabatic media so that all quantities can be scaled. We may thus consider whether our simulations represent both a quite compact radio galaxy and a giant source. For the example parameters detailed in Table~\ref{parameters}, the scale size and dynamical time scales are 75~kpc and 18~Myr (Compact) and 750~kpc and 148~Myr (Giant)

\begin{table*}
\caption{Breakdown of simulation names and main parameters that are used in this paper. "z" in the file name signifies a reflective inflow boundary}
\label{simulation_name}
\begin{tabular}{ccccccc}
\hline
Code Used & File Name & Directory Name & Resolution & Density Ratio & Mach & Precession $\frac{\pi}{180}$(rad) \\ 
\hline
Both & zaa & xyz1 & 75x75x75 & 0.1 & 6 & {1}   \\
 \hline
Both & ba & xyz1 & 150x150x150 & 0.1 & 6 & {1}   \\
Both & bb & xyz01 & 150x150x150 & 0.01 & 6 & {1}   \\
Both & bc & xyz001 & 150x150x150 & 0.001 & 6 & {1}  \\
Both & bd & 10xyz1 & 150x150x150 & 0.1 & 6 & {10}  \\
Both & bg & 20xyz1 & 150x150x150 & 0.1 & 6 & {20}  \\
Both & bm & xyz0001 & 150x150x150 & 0.001 & 6 & {1}  \\
Both & bn & 10xyz0001 & 150x150x150 & 0.1 & 6 & {10}  \\
Both & bo & 20xyz0001 & 150x150x150 & 0.1 & 6 & {20}  \\
Both & zbm & Rxyz0001 & 150x150x150 & 0.001 & 6 & {1}  \\
Both & zbn & 10Rxyz0001 & 150x150x150 & 0.1 & 6 & {10}  \\
Both & zbo & 20Rxyz0001 & 150x150x150 & 0.1 & 6 & {20}  \\
Both & zba & Rxyz1 & 150x150x150 & 0.1 & 6 & {1}  \\
 \hline
Both & zca & Rxyz1 & 225x225x225 & 0.1 & 6 & {1}  \\ 
\hline
Both & zda & Rxyz1 & 300x300x300 & 0.1 & 6 & {1}  \\ 
\hline
PLUTO CODE & ea & xyz1 & 150x150x150 & 0.1 & 2 & {1}  \\
PLUTO CODE & ed & 10xyz1 & 150x150x150 & 0.1 & 2 & {10}  \\
PLUTO CODE & ee & 20xyz1 & 150x150x150 & 0.1 & 2 & {20}  \\
PLUTO CODE & fa & xyz1 & 150x150x150 & 0.1 & 4 & {1}  \\
PLUTO CODE & fd & 10xyz1 & 150x150x150 & 0.1 & 4 & {10}  \\
PLUTO CODE & fe & 20xyz1 & 150x150x150 & 0.1 & 4 & {20}  \\
PLUTO CODE & ga & xyz1 & 150x150x150 & 0.1 & 8 & {1}  \\
PLUTO CODE & ha & xyz1 & 150x150x150 & 0.1 & 12 & {1}  \\
PLUTO CODE & ia & xyz1 & 150x150x150 & 0.1 & 24 & {1}  \\
PLUTO CODE & id & 10xyz1 & 150x150x150 & 0.1 & 24 & {10}  \\
PLUTO CODE & ie & 20xyz1 & 150x150x150 & 0.1 & 24 & {20}  \\
PLUTO CODE & ja & xyz1 & 150x150x150 & 0.1 & 48 & {1}  \\
\hline
\end{tabular}
\end{table*}
 
The ambient medium parameters are specified through the number density and temperature, both constrained from X-ray data. We specify the hydrogen nuclei density here and add on 10\% of helium nuclei, assuming both species to be fully ionised.
  
In this work, we dump the data every 0.1 units, but this is altered according to the rate of propagation of the jet across the grid. This translates to 1 dump file roughly every 0.36~Myrs (Compact Source) and 3.0~Myr (Giant Source).

A major aim is to investigate the influence of jet  precession on the morphology of the radio galaxy. The precession is added into the system by splitting the velocity up into components. To achieve this, the full jet speed  is taken to be 
\begin{equation}\label{eq_initvelocity}
    v_\text{jet} =  M_\text{jet} \cdot{c}_\text{jet} = M_\text{jet}\cdot \sqrt{{{p}_\text{jet}}\over{{p}_\text{amb}}} \cdot \sqrt{{\rho _\text{amb}}\over{\rho _\text{jet}}} \cdot {c}_\text{amb},
\end{equation}
where $p_{\text{jet}}$, $\rho_{\text{jet}}$ and $M_\text{jet}$ are the pressure, density and Mach number of the jet, respectively. 
The jet speed is calculated using Mach numbers of 2, 4, 6, 8, 12, 24 and 48. 

The velocity components are then
\begin{align}\label{eq_velocity}
     &v_{\text{y}}=  v_{\text{jet}} . \sin(\theta) . \sin(\omega_{\text{l}} . t) \nonumber \\ 
     &v_{\text{z}}=  v_{\text{jet}} . \sin(\theta) . \cos(\omega_{\text{l}} . t) \\
     &v_{\text{x}}= \sqrt{{v_{\text{jet}}}^{2} - ({v_{\text{y}}}^{2} + {v_{\text{z}}}^{2})} \nonumber
\end{align}

The precession is added at the nozzle of the jet as it enters the inflow boundary. As seen in Equation (\ref{eq_velocity}), the angle of precession is $\theta$. To cover a wide range of scenarios, precession angles of 0.25\degsy, 1\degsy, 5\degsy, 10\degsy and  20\degsy are used. The smallest precession of 0.25\degsy is applied to break up any numerical fluctuation when we simulate a straight jet. 
By default precession of 1\degsy, 10\degsy and  20\degsy are used for all densities; the precession of 0.25\degsy and 5\degsy are for a more in-depth look for the particular case of the density ratio of 0.1. 
The rate at which the jets precess is set to a default of once per four time units. 

\subsection{Mass, momentum and energy}

The mass flux injected into the system is 
\begin{equation}\label{eq_mass_flux}
      \dot {\mathcal{M}}_\text{jet} =  \rho_\text{jet}  \cdot  v_x  \cdot A,
\end{equation}
where $\rho_\text{jet}$ is the input jet density, $v_x$ is the jet velocity normal to the boundary, which is obtained from Equation (\ref{eq_initvelocity}), and 
$A =  (1-\mu)\pi r_\text{jet}^2$ is the jet area. Here, $\mu$ represents a small adjustment since the numerical nozzle profile is an approximation to a circle. There will be a linear increase with time for all the simulations with reflective outflow boundaries because the mass influx is constant throughout the present set of simulations. With an outflow boundary condition, the cocoon back flow removes mass from the grid, as discussed below.

In order to test and calibrate the results, we introduce the  Steady Propagation Model. The model assumes that the jet ploughs into the ambient medium, advancing a high-pressure hot spot at a constant speed. To understand how the jet and lobe propagate through the ambient medium, we also assume here that the jet density is low and the Mach number is high. We then write the jet momentum flow rate along the axis in terms of the ram force
\begin{equation}\label{eq_mass_flux}
      \text{\.{P}} =  \rho_\text{jet}  \cdot  v_x^2  \cdot A.
\end{equation}
This can be written in the form
\begin{equation}
     \text{\.{P}} =   \rho_\text{jet} \cdot  \frac{v_x^2}{v_\text{jet}^2}  \cdot M^2 \cdot c_\text{jet}^2 \cdot  A.
\label{eq-momentum}
\end{equation}
In this model, we take $p_\text{jet} = p_\text{amb}$ so that 
\begin{equation}\label{eq_mass_flux}
    \text{\.{P}} =  \rho_\text{amb}  \cdot c_\text{amb}^2 \cdot  \frac{v_x^2}{v_{jet}^2}  \cdot M_\text{jet}^2 \cdot A.
\end{equation}
That means that the momentum flow rate is roughly a constant for a fixed Mach number, independent of the jet density.
Remarkably, this implies that the jet crossing time of all low-precession simulations of a given size and Mach number  is a constant.
The speed scale, $U$ for the advance of the radio galaxy is given by the momentum balance formula,
\begin{equation}\label{eq_mass_flux}
     U^2 =  \frac{\rho_\text{jet}}{\rho_\text{amb}}  \cdot  v_x^2 = M_\text{jet}^2 \cdot c_\text{amb}^2,
 \end{equation}
which assumes that the jet remains collimated and propagates with a drag coefficient of unity. This yields a radio galaxy crossing time of 
\begin{equation}\label{eq_mass_flux}
      D/U = D/(M_\text{jet} \cdot c_\text{amb}).
\end{equation}
Therefore, we expect the jet crossing time to be approximately 5 units for the $D = 30$ grid and $ M = 6$ jet.
 
A cocoon can usually be easily identified with a radio galaxy simulation. This is filled with the jet material which has been shocked at the hot spot and spills out into a lobe or cocoon immediately surrounding the jet. The cocoon is thus distinct from the shocked ambient material, as originally defined \citep{1982A&A...113..285N}, but does not include the shocked ambient (as taken by
\citet{1992ApJ...392..458C}).

If the cocoon is cylindrical, the volume then grows as $\pi (R_c^2-r_\text{jet}^2) D$ where $R_c(t)$ is the root-mean-square average cocoon radius. We also assume that the jet flow is braked at the hot spot by a strong shock with a compression ratio of $(\gamma+1)/(\gamma-1) =  4$ and a hot-spot pressure of $2{\gamma}M_\text{jet}^2p_\text{jet}/({\gamma}+1)$. This is followed by an adiabatic pressure fall back to the ambient/jet pressure. The associated density decrease directly yields the cocoon density as
\begin{equation}\label{eq_cocoon1}
        \rho_{c} =    \frac{\gamma+1}{\gamma-1}  \left[ \frac{\gamma+1}{2\gamma M_\text{jet}^2 }\right]^{1/\gamma} \rho_\text{jet}, 
   \end{equation}  
   or, for $\gamma = 5/3$,  
   \begin{equation}\label{eq_cocoon2}          
        \rho_{c} =              4 \left({\frac{4}{5}}\right)^{3/5} M_\text{jet}^{-6/5} \rho_\text{jet}.
 \end{equation} 
 
 Under a reflection boundary condition, we equate the injected mass to the cocoon mass,  which then yields the cocoon volume and 
 manipulation then yields the average cocoon radius through
 \begin{equation}
       R_c =  r_\text{jet}  \left[ 1 +  \frac{5^{3/5}}{4^{8/5}} { M_\text{jet}^{6/5} ( \frac{\rho_\text{amb}}{\rho_\text{jet}}})^{1/2}  \right]^{1/2}.
 \label{cocoonradius}
 \end{equation} 
It can thus be seen that even with jet-ambient density ratios of between 0.1 and 0.0001, a Mach 6 jet would generate an averaged cylindrical lobe of radius of between 3.0 and 15.7~$r_\text{jet}$.
 
The assumed pressure equilibrium with the ambient medium, however, cannot be established if the cocoon is too wide, occupying the region which the shocked ambient medium would have expanded into. Instead the shocked ambient medium does not re-expand and so applies a surface pressure on the cocoon. Equation~\ref{cocoonradius} implies that this will occur at very low jet densities or extremely high Mach numbers.
  
The energy pumped on to the grid is converted or diverted into several components. The added energy is contained in the ambient medium, cocoon or jet. For each, we have contributions to both  the thermal and kinetic (turbulent) energy. To obey conservation of energy, we need to account for any energy lost through the boundaries, although this is zero provided no disturbances have reached the outer boundaries and we have  imposed a reflection inner boundary condition. The total power added to the grid is:
 
\begin{equation}\label{eq_energy_flux}
      \text{L} =   \left( \frac{1}{2}v_\text{jet}^2 +  \frac{1}{\gamma-1}\frac{p_\text{jet}}{\rho_\text{jet}}\right) \cdot  v_x\cdot \rho_\text{jet} \cdot A,
\end{equation}
which, on taking $v_x = v_\text{jet}$ for simplicity,  can be written
\begin{equation}\label{eq_energy_flux}
      \text{L} =   \left( \frac{1}{2} +  \frac{1}{\gamma} \frac{1}{\gamma-1} \frac{1}{M_\text{jet}^2}\right)  \cdot 
      \left( \frac{\rho_\text{amb}} {\rho_\text{jet}} \right)^{1/2} 
      \cdot  M_\text{jet}^3  \cdot  \rho_\text{amb} \cdot   c_\text{amb}^3 \cdot A.
\end{equation}
Again, note the low dependence on the density ratio and the strong dependence on Mach number. In Section \ref{energy}, we will determine how this energy is redistributed.

Radio  maps and X-ray images will be derived from these simulations in a following work. Here, we utilise pseudo-synchrotron radio emission by taking the emission per unit zone volume as
\begin{equation}\label{eq_sudo_synchrotron}
E_{\text{radio}} \propto  \chi \cdot p^{2},
\end{equation}
where the tracer $\chi$ is set so that the ambient has a value of 0 and 100\% jet is set to a value of 1, thus only selecting zones where  there is material originating from jet injection but not accounting for shock acceleration. By summing the emissivity through a specific direction we obtain emission maps and can thus find the location of the maximum intensity which we use below as a quantitative measure of the  numerical convergence. 

A list of simulations and their naming conventions is provided in Table~\ref{simulation_name}. There are two main names for each jet simulation. The first is the file name that ZEUS-3D uses which is carried over to the PLUTO code. The second is comprised of three parts [precession][coordinate system][density ratio]. Also, if there is a reflective boundary added to the simulation then an ``R" is added to the corresponding coordinate. For example, 20Rxyz01 refers to  20$^\circ$ precession in Cartesian coordinate with a reflection in the starting plane of the jet with a density ratio of 0.01. Note that the  reflection condition is  only applied to the boundary where the jet is initialized. 

\section{Resolution \& Convergence}
\label{resolution}

Resolutions studied span the range from $75^3$ to $300^3$. Our workhorse is the $150\,^{3}$ simulation. This permits us to cover as much ground as possible and follow up with higher resolutions on any interesting findings. The wide range allows us to test the influence of the resolution on the morphology of the jet. All the simulation which are illustrated have the conditions outlined in Table~\ref{simulation_name}.  

\begin{figure*} 
\begin{center}$
\begin{array}{cc}
  \includegraphics[width=0.5\textwidth]{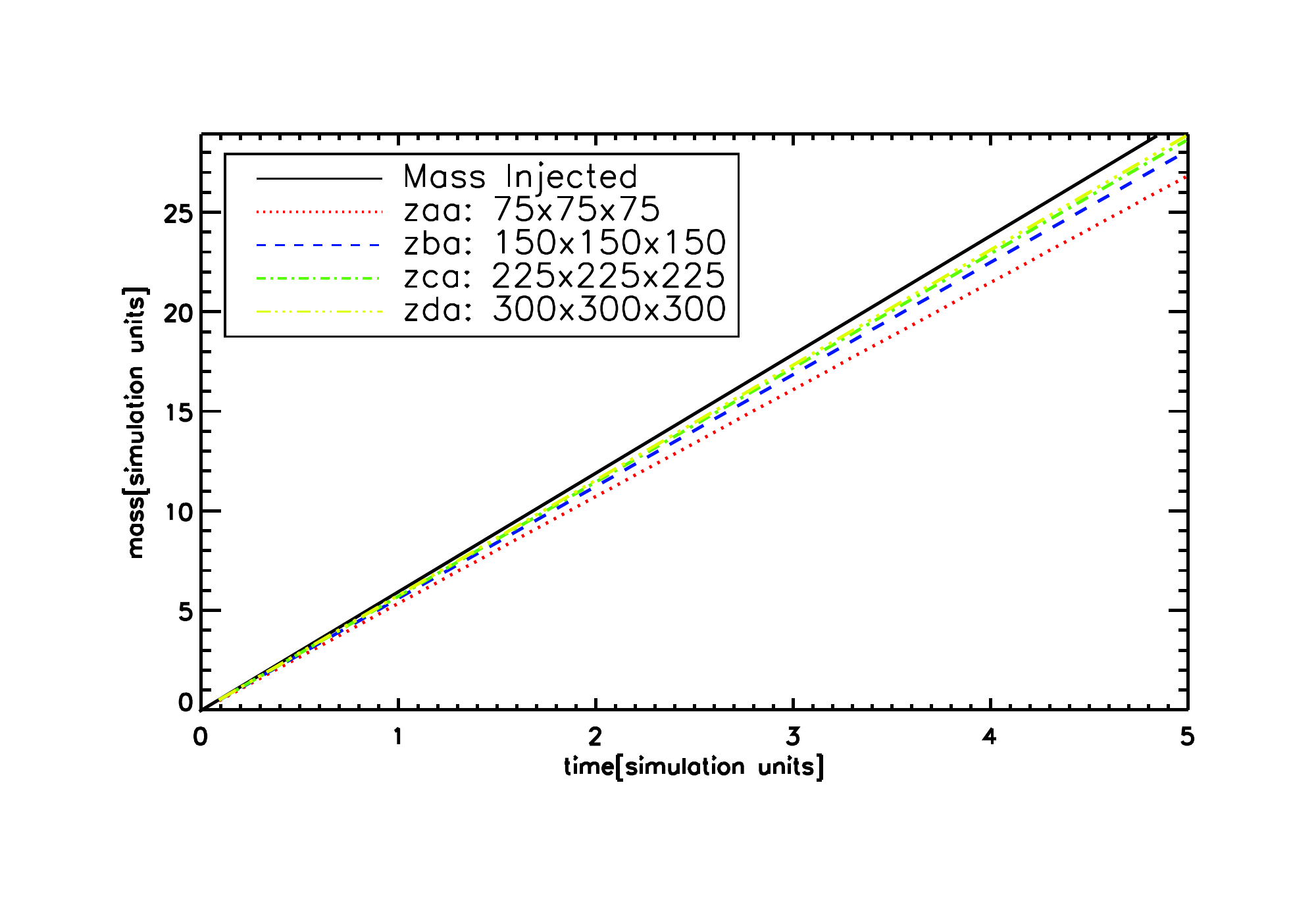} &
  \includegraphics[width=0.5\textwidth]{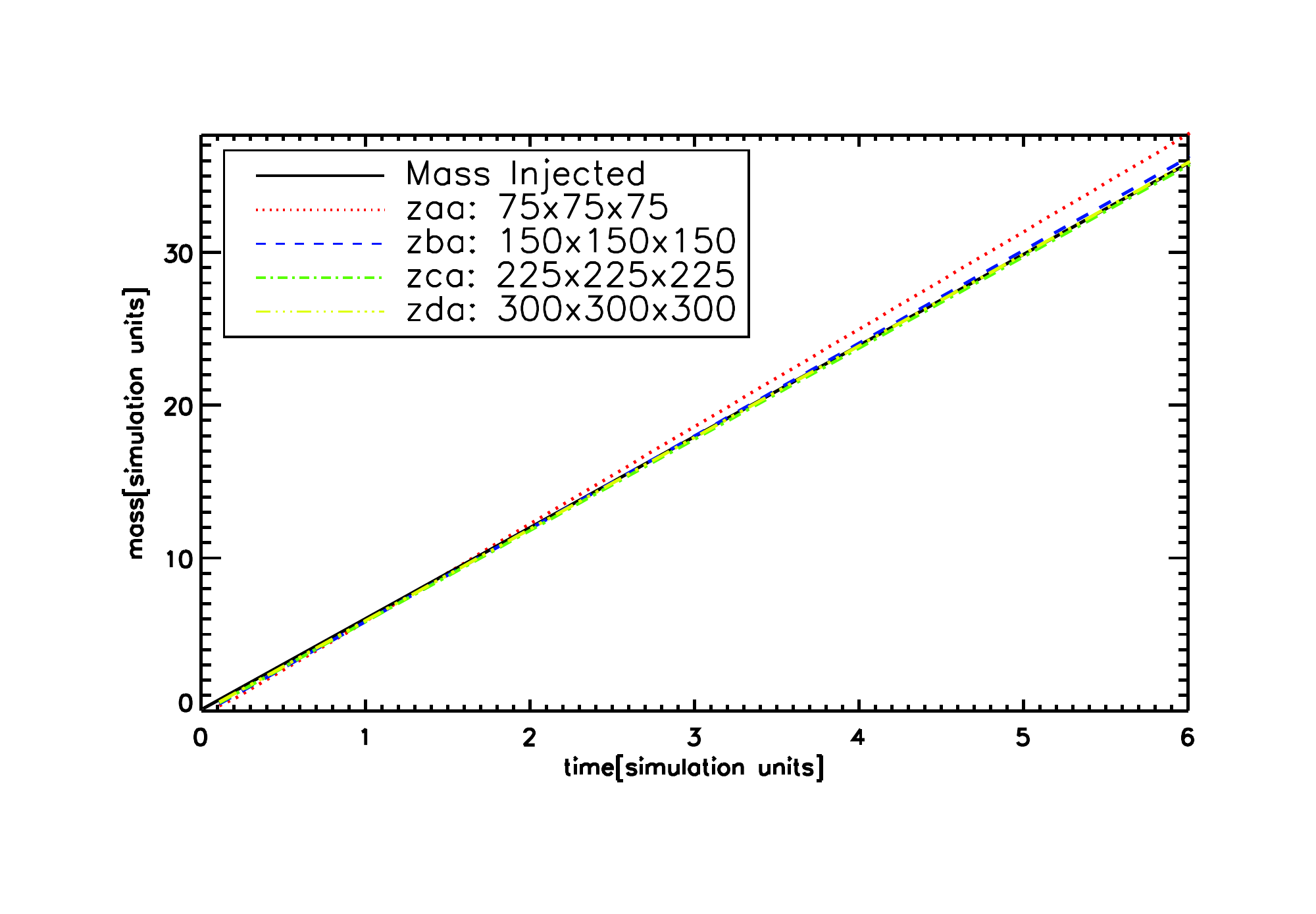}
\end{array}$
\end{center}
\caption{The total mass on the grid as a function of time for the reflection boundary condition on the inflow boundary, generated for the indicated four resolutions.  The density ratio is $ \rho_\text{jet}/\rho_\text{amb} = 0.1$ and the precession angle is  1$^\circ$. The left panel displays ZEUS-3D and the  right shows the PLUTO code. The solid black line corresponds to the theoretical value that should be injected into the system through a perfect circular cross-section. }
\label{dz_p_reflective_mass_resolution_graph}
\end{figure*}

The accumulated mass injected as a function of time is displayed in Fig.~\ref{dz_p_reflective_mass_resolution_graph}. It is calculated from the grid density and tracer dumps with the reflection boundary condition applied.
The tracer provides an average for zones where mixing occurs with the ambient medium while ambient mass loss from the grid at late times
is discounted. Jet material does not leave the grid since the reflection boundary condition is applied across the entire inner plane upon which the jet nozzle is then superimposed. 

Figure~\ref{dz_p_reflective_mass_resolution_graph}
shows that there is a clear discrepancy with the theoretical mass inflow of 5.96 per unit time assuming a perfect circular nozzle of radius $r_\text{jet}$ as shown by the solid lines. The cause of the discrepancy is made clear from the dependence on the resolution. As the resolution is increased, the mass inflow rate converges towards a value within a few per cent of the theoretical value for the PLUTO code (right panel). However, this value is lower for the ZEUS runs (left panel): the ZEUS-3D runs converge under the theoretical line whereas the PLUTO code is converging onto the theoretical line. In both cases, we can see the injected mass starts to converge at a resolution of 150$^{3}$. 

Each code sets up the circular jet nozzle on the square faces of the entry zones by applying a different approximation scheme. A smoothing profile is used to fix the  speed and mass mixing in the interface zones. The resolution thus influences the mass inflow according to the number of zones across the jet diameter $N/D$ where $N$ is the grid resolution. Fig.~\ref{dz_p_reflective_mass_resolution_graph} shows that the mass discrepancy is approximately $\propto (D/N)^2$. This result highlights the fact that jet simulations are extremely sensitive to the frayed edges of the jet surface which may introduce instabilities and small-scale turbulence on larger scales much further downstream. For this reason, we do not expect any two codes, let alone any two resolutions with the same code, to generate  exactly the same physical  structures. However, some  of this sensitivity is clearly eliminated at higher resolutions.

\begin{figure*}
\begin{center}$
\begin{array}{cc}
  \includegraphics[width=0.5\textwidth]{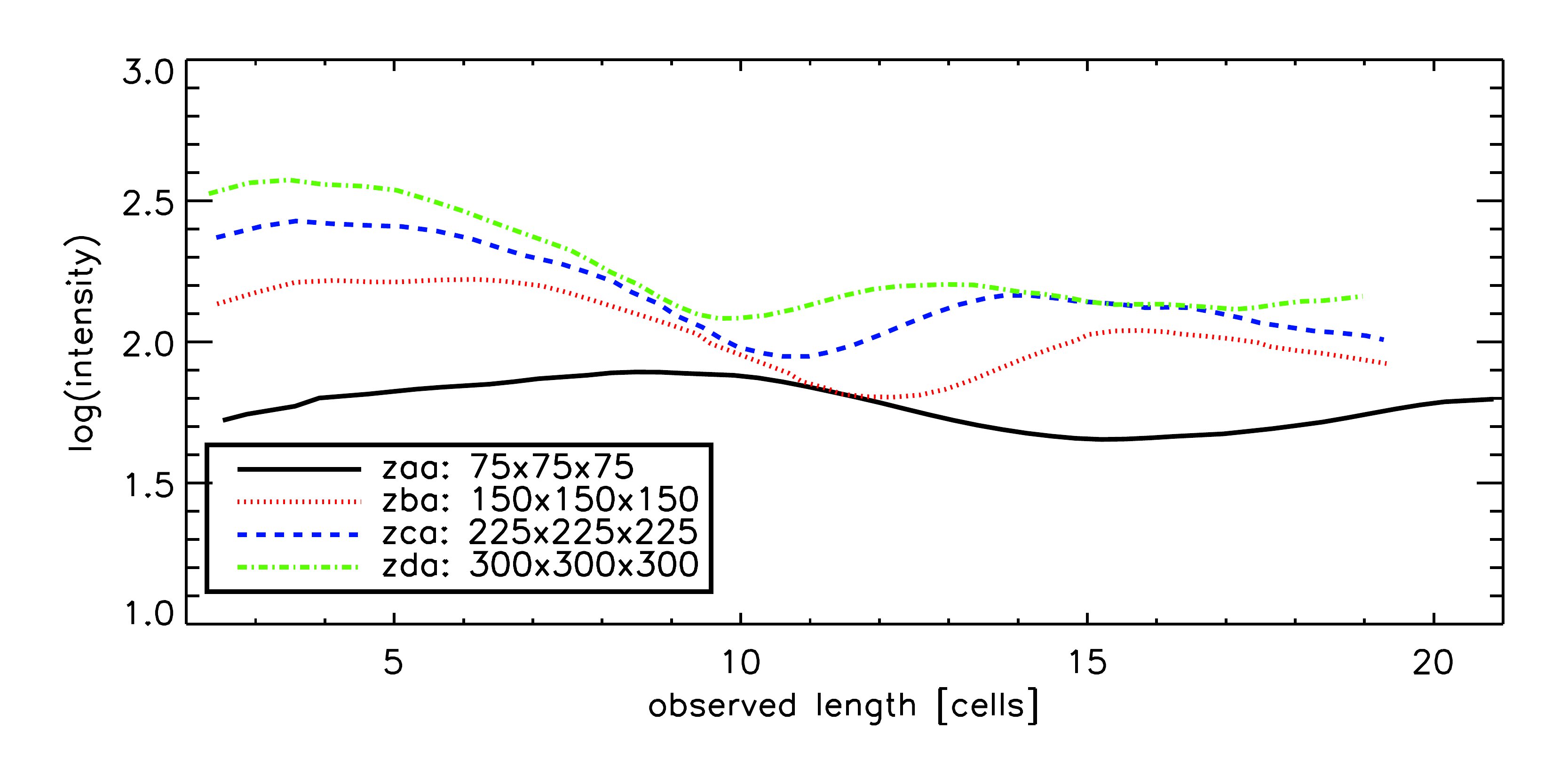} &
  \includegraphics[width=0.5\textwidth]{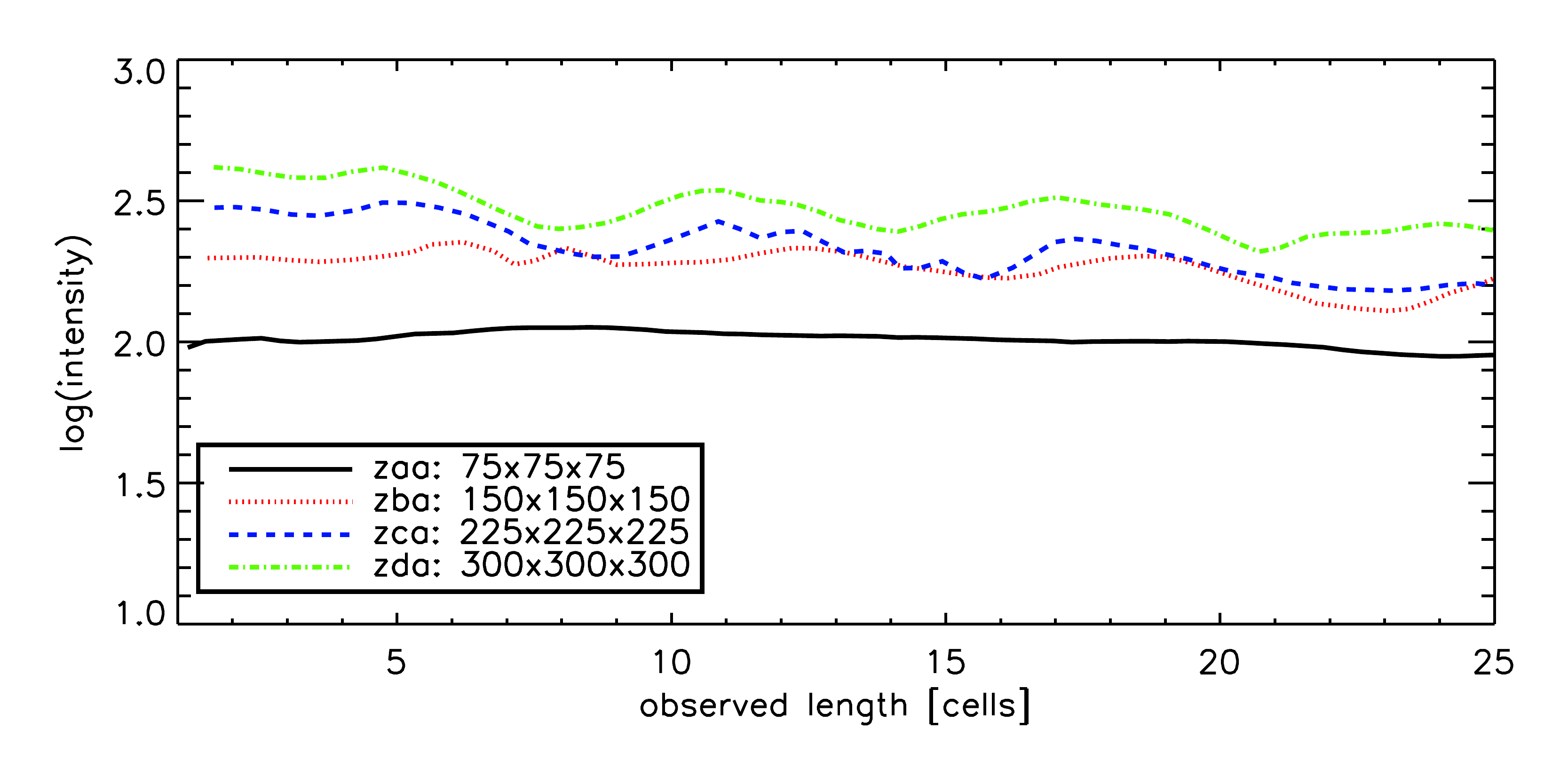}
\end{array}$
\end{center}
\caption{{\small The maximum intensity of the hot spot and the distance of that hot spot to the source.  Left graph shows ZEUS-3D; right shows PLUTO results. These tests were performed with a reflection inflow boundary.}}
\label{dz_p_reflective_intensity_resolution_graph}
\end{figure*}

In the low resolution runs, the lack of detail and the averaging of the eddy currents that occur at the interface of the jet and ambient material are clear. As one way to  quantify this, we  study the maximum intensity of the ``hot spot" on the pseudo-synchrotron images, and the distance of that hot spot from the source as a function of the resolution. This is a surface brightness and, hence, would be expected to increase gradually as the resolution increases. This is displayed in Figure~\ref{dz_p_reflective_intensity_resolution_graph} for ZEUS-3D and PLUTO.  There are considerable variations as expected although these variations decrease as the jet radius becomes better resolved. There is an indication that PLUTO starts to converge at resolution greater than 225$^3$. 

\begin{figure*}
\begin{center}$
\begin{array}{cc}
  \includegraphics[width=0.5\textwidth]{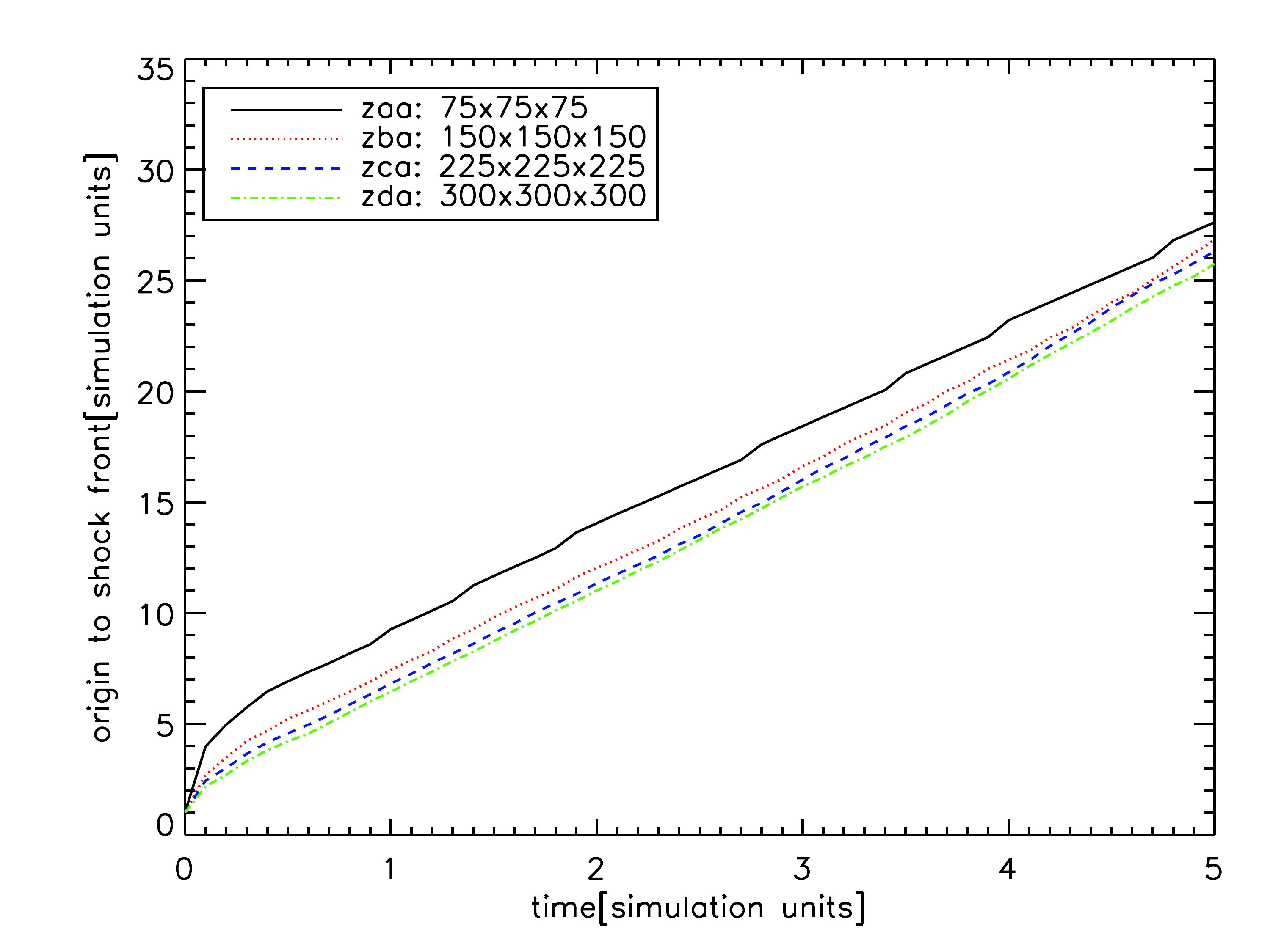} &
  \includegraphics[width=0.5\textwidth]{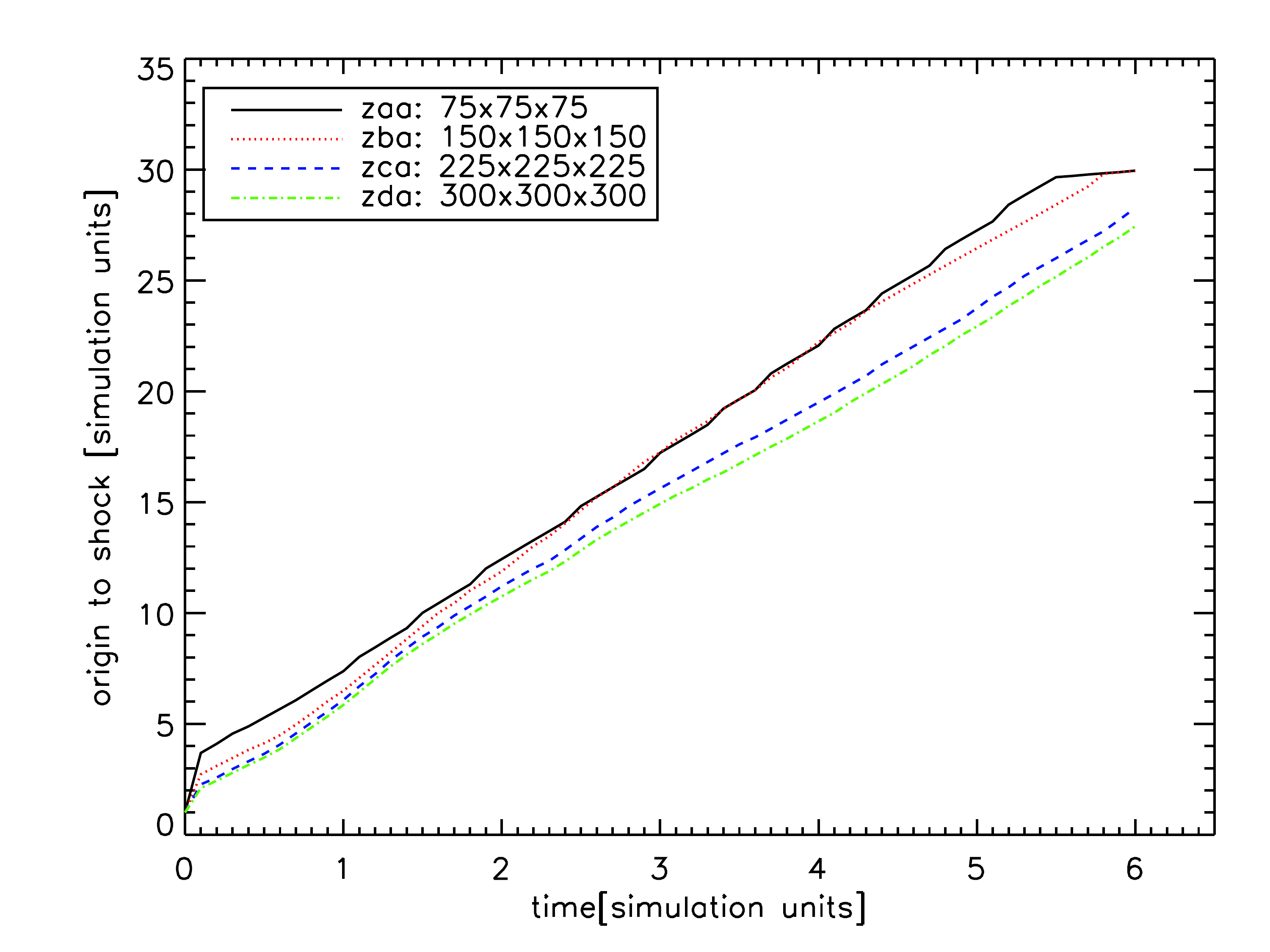}
\end{array}$
\end{center}
\caption{{\small  A resolution study of the location of the advancing shock into the ambient medium for the case with jet-ambient density ratio of  0.1 and outflow boundary condition. The resolutions shown increase from 75$^3$ (solid, black) to 300$^3$ (dot-dash, green) with the initial conditions outllned in
 Table~\ref{simulation_name}.  The left panel  is ZEUS-3D code and the right panel is the PLUTO code.}}
\label{dz_p_shock_propagation_resolution_graph}
\end{figure*}

This is further supported when the advance of the ambient shock front is plotted, as shown in Fig.~\ref{dz_p_shock_propagation_resolution_graph}. The advance speed is stable between the resolution runs after an initial set-up period during which the speed of the abrupt entrance depends on the grid zone size as the code smooths over the steep gradients introduced. The advance speed settles to a constant value of $\sim$~4.7 and 4.3 for the ZEUS-3D and PLUTO codes, respectively. This is lower than the value of $ U = 6$ for the Steady Propagation Model. However, as will be found below, advance speeds in excess of $U$ are found for lower jet densities and, in contrast,  can also be higher for the PLUTO code. In summary, with specific reservations, the 150$^3$ resolution is the preferred choice for a full numerical survey.

\section{Parameter Study}
\label{parameter}

\subsection{Density}

\begin{figure*} 
\includegraphics[width=0.9\textwidth]{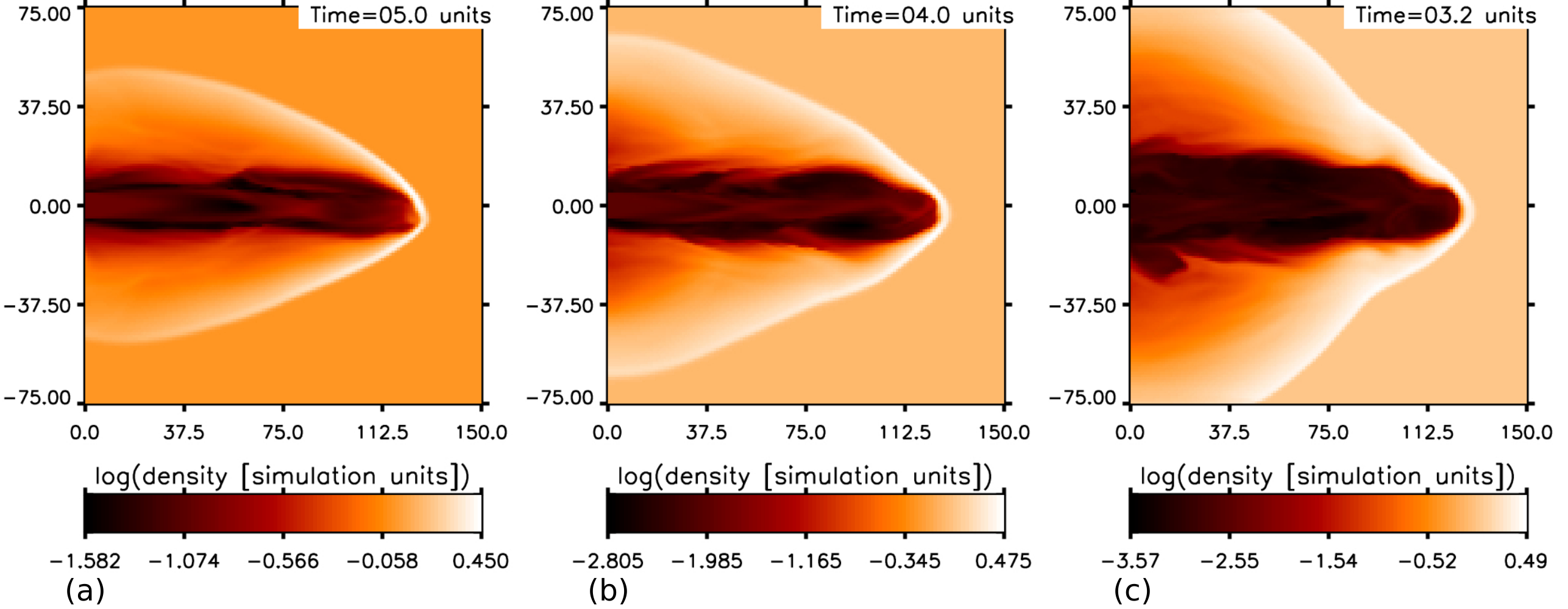}
\caption{{\small Density dependence. Density volumetric slices through the mid-plane from ZEUS-3D with equivalent conditions apart from the jet-ambient density ratio of  0.1 (a),  0.01 (b) and 0.001 (c). All have a jet Mach number of 6, a small precession angle of 1$^\circ$ and period of 4 simulation units in order to break the symmetry. Conditions are outlined in Table~\ref{simulation_name}. The limits of the colour scale are set to the specific simulation minimum and maximum values.}}
\label{dz_rho_xyz1_01_001}
\end{figure*} 

\begin{figure*} 
\includegraphics[width=0.9\textwidth]{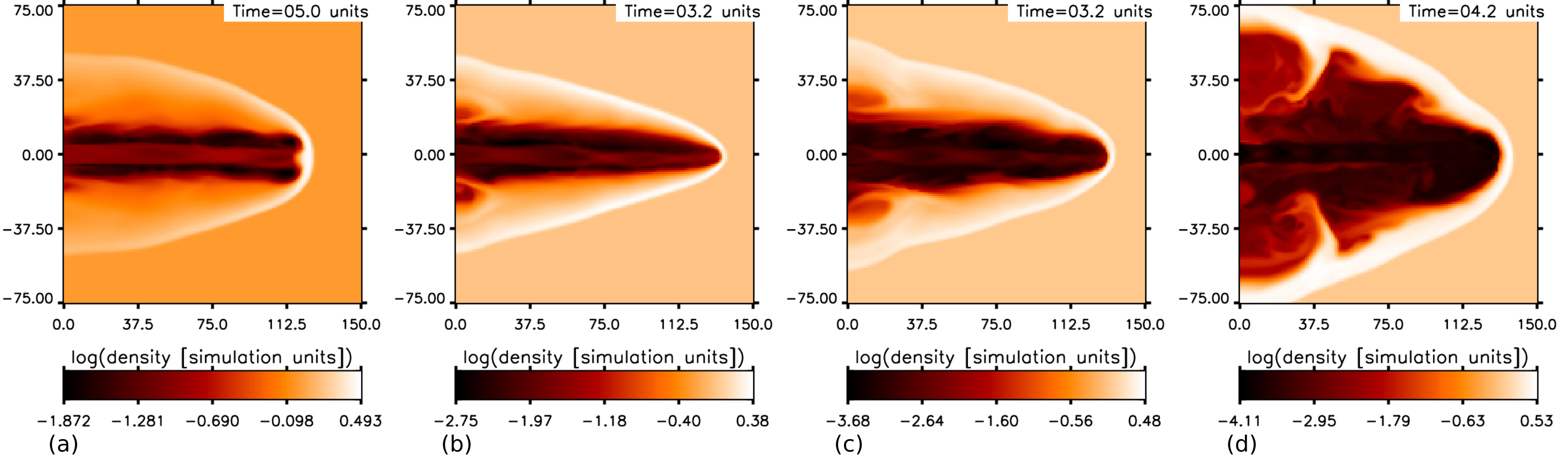}
\caption{{\small Density dependence. Density volumetric slices from the PLUTO code with equivalent conditions apart from the jet-ambient density ratio of 0.1 (a), 0.01 (b), 0.001 (c) and 0.0001 (d). Conditions are as stated in Fig.~\ref{dz_rho_xyz1_01_001}.}}
\label{p_rho_xyz1_01_001_0001}
\end{figure*} 

\begin{figure*}
\includegraphics[width=0.9\textwidth]{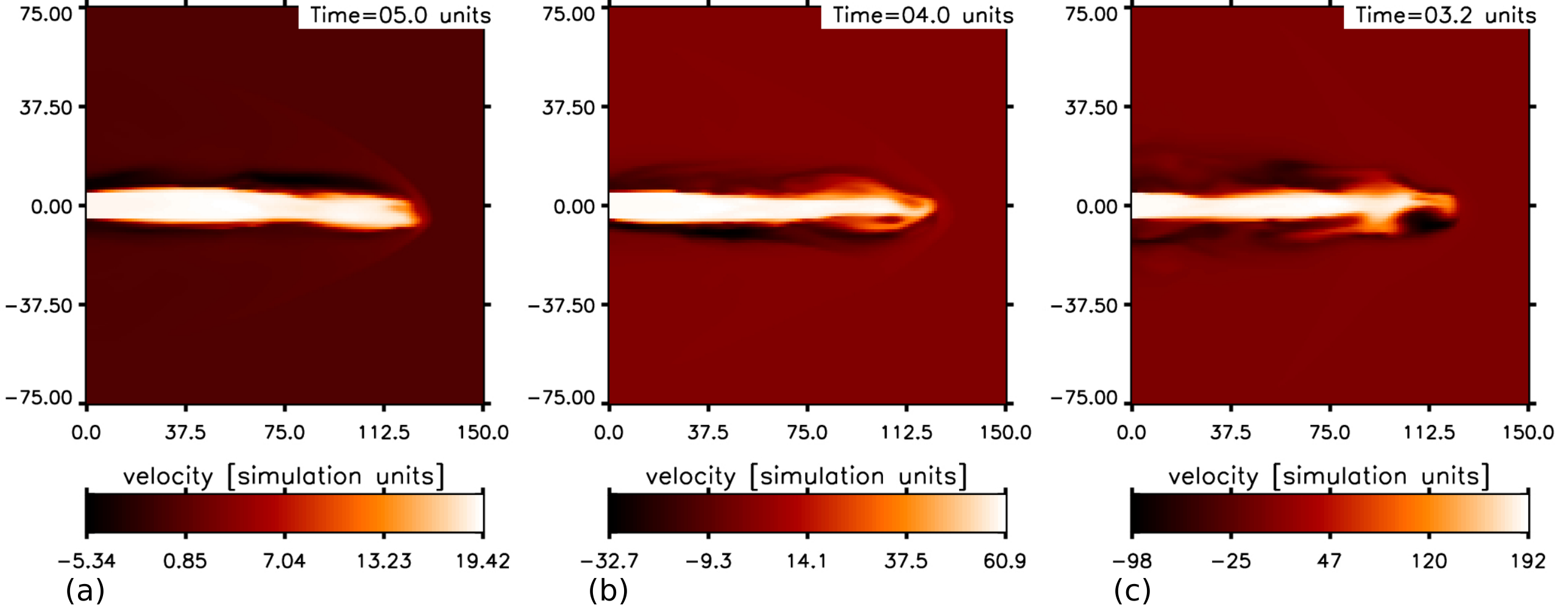}
\caption{{\small Velocity volumetric slices from ZEUS-3D with equivalent conditions bar from the density ratio of 0.1 (a), 0.01 (b) and 0.001 (c).
Other conditions are as stated in Fig.~\ref{dz_rho_xyz1_01_001}.}}
\label{dz_vel_xyz1_01_001}
\end{figure*}

\begin{figure*}
\includegraphics[width=0.9\textwidth]{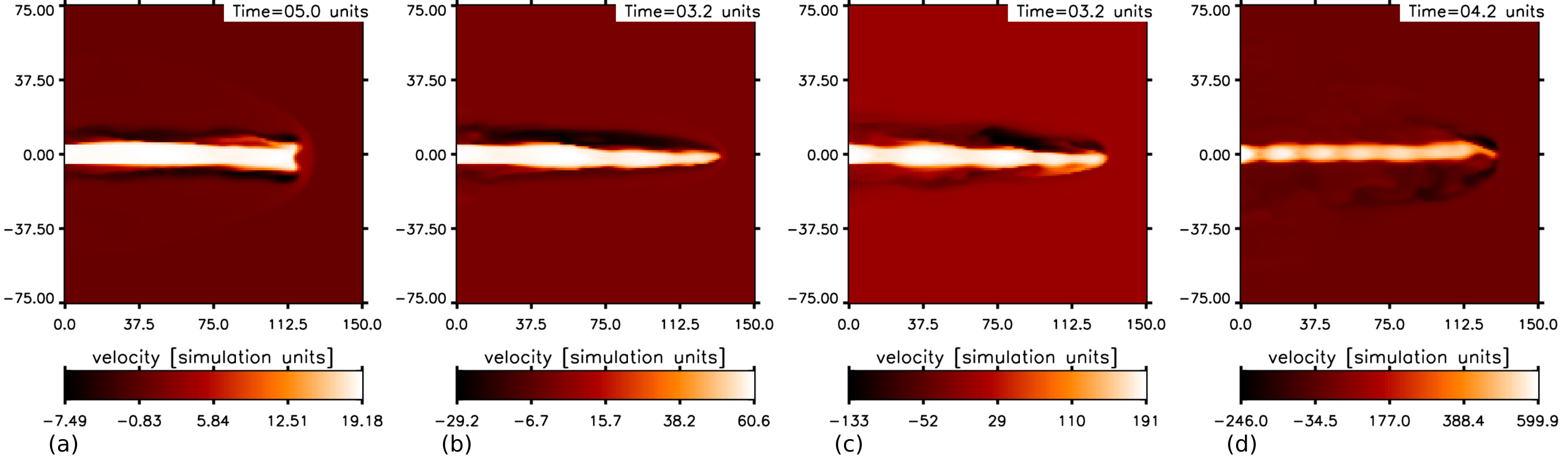}
\caption{{\small Velocity volumetric slices from PLUTO CODE with equivalent conditions bar from the density ratio of 0.1 (a), 0.01 (b), 0.001 (c) and 0.0001 (d). Other conditions are as stated in Fig.~\ref{dz_rho_xyz1_01_001}.}}
\label{p_vel_xyz1_01_001_0001}
\end{figure*}

The dependence on the jet-ambient density ratio  is illustrated in  the density slices displayed in Figs.~ \ref{dz_rho_xyz1_01_001} \& \ref{p_rho_xyz1_01_001_0001} and the corresponding velocity  slices of Figs.~ \ref{dz_vel_xyz1_01_001} \& \ref{p_vel_xyz1_01_001_0001}.  These correspond to straight collimated jets with a superimposed small-angle long-period precession to break the symmetry. 

The density slices show that there is a modest increase in the volume of the cocoon occupied by jet material  as the density ratio decreases below 0.01.
The  bow shock in the ambient medium also becomes progressively blunter as the jet density falls.  

The narrow cocoon associated with density ratios above 0.001 ensures that the jet propagates with little dissipation across the jet-cocoon vortex sheet. In contrast, at lower densities,  the cocoon broadening is associated  with the generation of large asymmetric vortices which finally dominate for the density ratio of 0.0001. 
This broad cocoon expansion for very light jets, typically to almost the same size as the bow shock, was first found in axisymmetric simulations of \citet{2003A&A...398..113K} for Mach numbers above approximately three. The wide cocoon supports a high pressure and strong pressure variations which feed back on to the jet. The jet axial velocity becomes limited to a narrower and convergent-divergent channel (see Figs.~ \ref{dz_vel_xyz1_01_001} and \ref{p_vel_xyz1_01_001_0001}) as the cocoon pressure dominates. 

The differences between the two codes are quite modest with the PLUTO simulations somewhat more aerodynamic in shape  and speed across the grid for a specific density ratio. This is consistent with the better smoothing profile across the nozzle interface of the PLUTO code. The difference between the two codes is apparent on comparison of slice (b) between Figs.~ \ref{dz_rho_xyz1_01_001} \& \ref{p_rho_xyz1_01_001_0001}.  There is a time difference of 0.8 units or 2.9~Myrs for the Compact Source despite  the same initial condition. It can be seen from the PLUTO figures that the jets themselves are more stable compared to the ZEUS code simulations. This means that the momentum is efficiently pushing the head of the jet into new ambient material rather than dissipating the energy to create a widening plume.

The advance speed increases by a significant factor as the density ratio is lowered from 0.1. But this trend is reversed for the very lowest density where the asymmetric structure leads to a spreading and partial disruption of the jet which slows down the advance. In comparison, from the Steady Propagation Model we would anticipate a constant speed of advance. The difference is associated with pressure feedback onto the jet which squeezes the jet as seen from the velocity structure in panels (b) and (c) of Figs.~ \ref{dz_vel_xyz1_01_001} \& \ref{p_vel_xyz1_01_001_0001}. 
The higher advance speed also tends to streamline the cocoon as material flowing into the cocoon does not have to expand so far laterally. A further consequence is that the cocoon-jet density ratio generally increases as the jet density decreases (see the associated colour bars).  

We thus confirm here that a wide range in jet density has a relatively small affect on the propagation speed of the lobes. This is because the injected momentum flow rates are the same. This is due to the jet density being inversely proportional to the sound speed squared of the jet. So decreasing the density ratio increases the velocity of the jet since the Mach number is held at a constant value, as seen from Equation \ref{eq_initvelocity}. This is also evident from the velocity slices displayed in Figs.~ \ref{dz_vel_xyz1_01_001} and \ref{p_vel_xyz1_01_001_0001}. The result is apparent when looking at the four density ratio simulations of PLUTO (Fig.~\ref{p_rho_xyz1_01_001_0001}). It shows that the time it takes for the density ratio of 0.1, 0.01, 0.001 and 0.0001 correspond to 18.1, 11.6, 11.9 and 15.2 Myr, respectively when interpreted as compact radio sources. Thus the momentum of the jet is the key factor in the propagation distance of the head of the jet. 

\begin{figure*}
\includegraphics[width=1.0\textwidth]{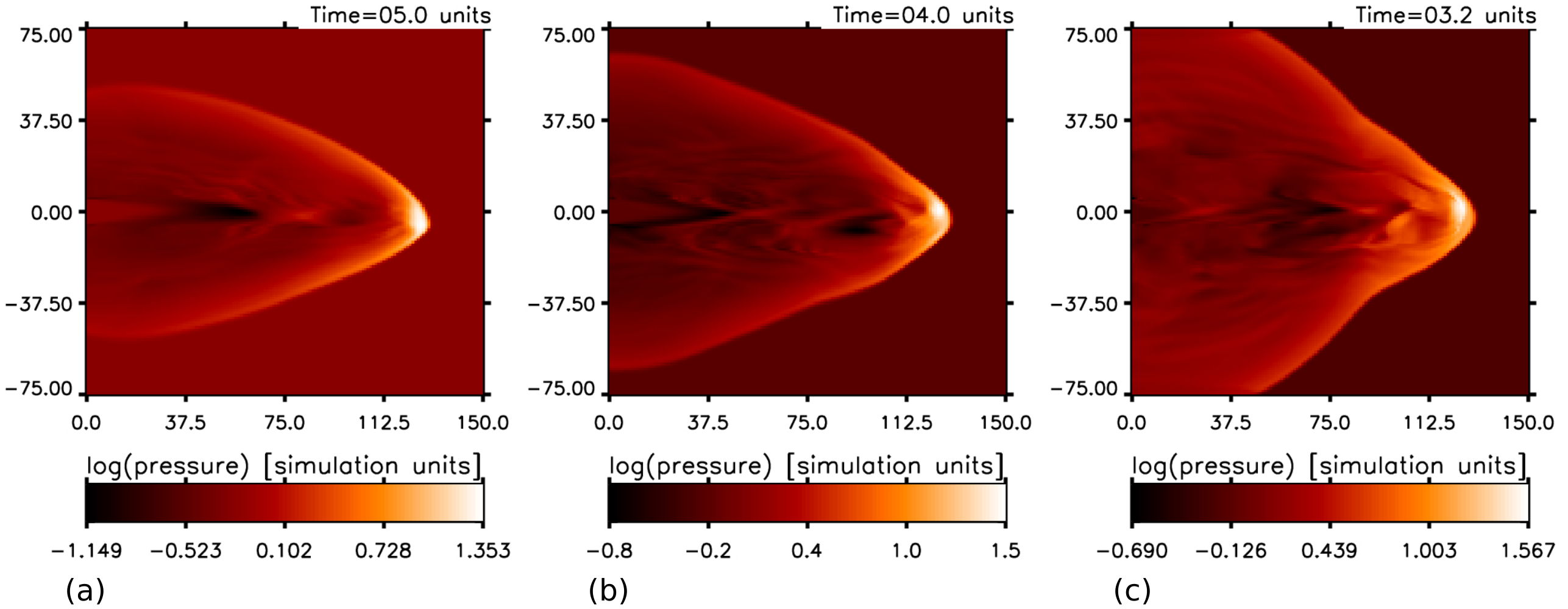} 
\caption{Pressure volumetric slices from ZEUS-3D with the conditions the same apart from the density ratio. Slice (a) shows a density ratio of 0.1, (b) is a ratio of 0.01 and (c) is a ratio of 0.001.  Other conditions are as stated in Fig.~\ref{dz_rho_xyz1_01_001}.  }
\label{dz_prs_xyz1_01_001}
\end{figure*}

\begin{figure*} 
\includegraphics[width=1.0\textwidth]{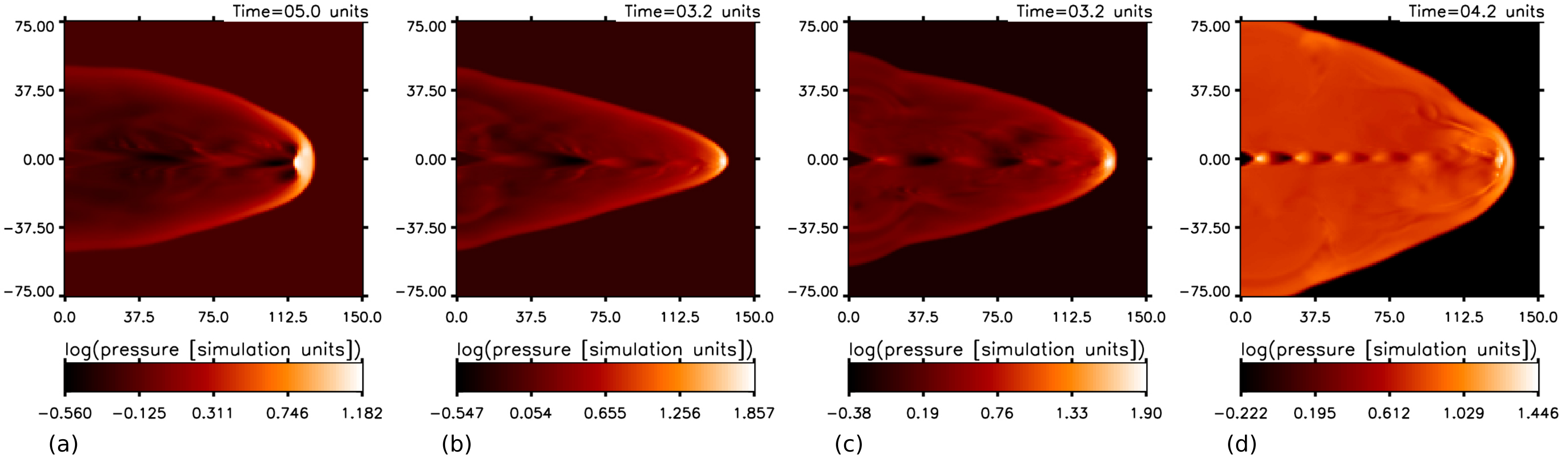}
\caption{Pressure volumetric slices from PLUTO with the conditions the same apart from the density ratio. Slice (a) shows a density ratio of 0.1, (b) is a ratio of 0.01 and (c) is a ratio of 0.001 and (d) is a ratio of 0.0001.  Other conditions are as stated in Fig.~\ref{dz_rho_xyz1_01_001}.   }
\label{p_prs_xyz1_01_001_0001}
\end{figure*}

The pressure distributions display more variety, as displayed in Figs.~\ref {dz_prs_xyz1_01_001} and \ref{p_prs_xyz1_01_001_0001}. While similar ambient bow shocks are present for the jet-ambient density ratio of 0.1 (left panels), the ambient bow is wider in the ZEUS runs although the maximum pressure reached is lower. Hence, the results are sensitive to the precise conditions with the PLUTO code generating slightly more  collimation and less
jet-cocoon turbulence at the lower densities.  Finally, the high pressure created at the lowest density (right panel) perturbs the under-pressured jet, leading to the strong pressure variations all along the jet, typical of a convergent-divergent nozzle.

\subsection{Parameter Study: Mach number}
\label{machnumber}

\begin{figure*}
\includegraphics[width=0.9\textwidth]{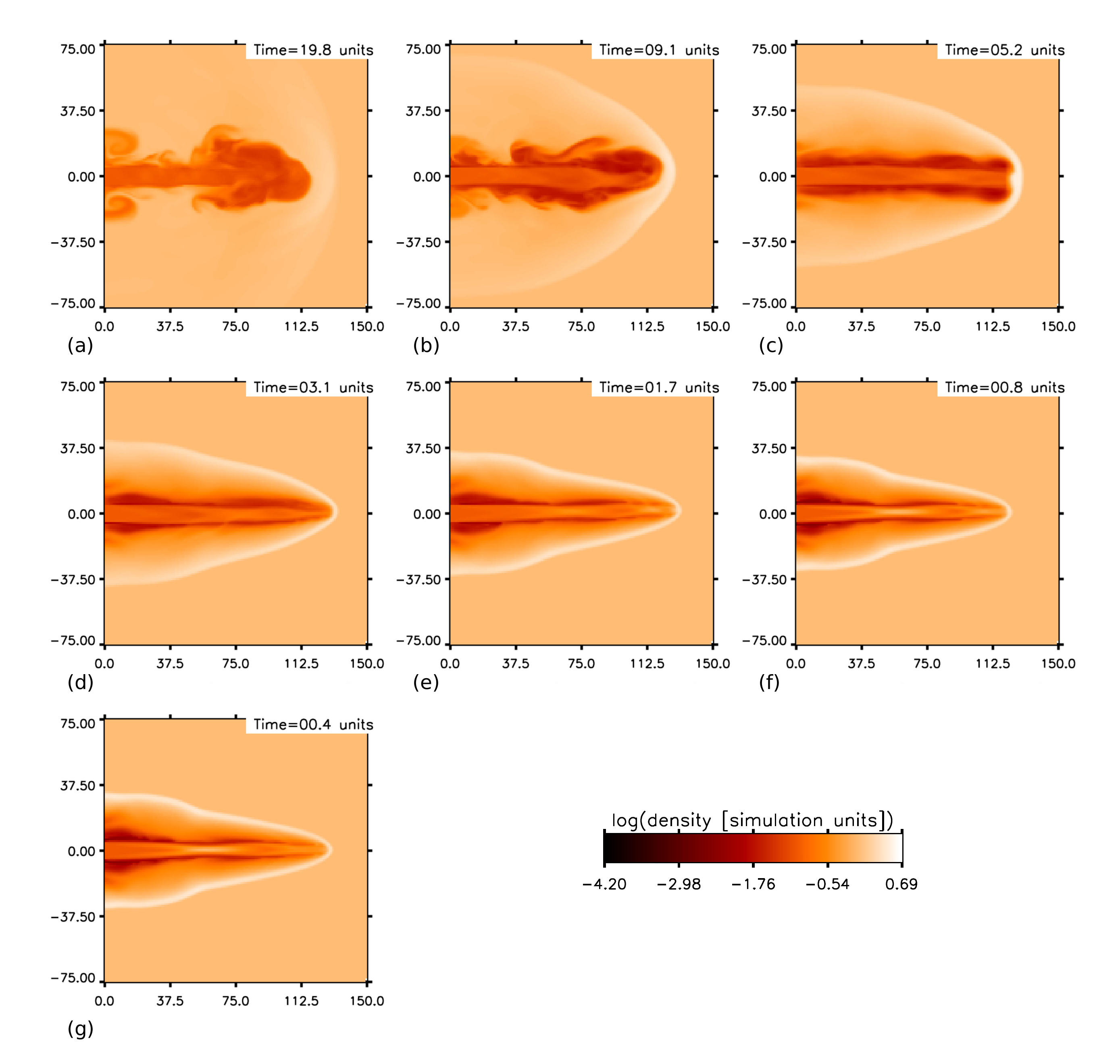}
\caption{{\small Mach number dependence. Density volumetric slices from the PLUTO code with equivalent conditions bar from Mach numbers of 2 (a), 4 (b), 6 (c), 8 (d), 12 (e), 24 (f) and 48 (g).  The same colour bar applies to  all slices. Other conditions are as stated in Fig.~\ref{dz_rho_xyz1_01_001}.}}
\label{p_M2_M4_M8_M12_M24_M48_rho_1}
\end{figure*}

For a fixed Mach number and pressure balance between the injected jet and ambient  medium, the momentum flow rate through the jet is independent of the jet density in the Steady Propagation Model (Eq.~\ref{eq-momentum}). However, the momentum flow rate through the jet is proportional to the Mach number squared. Therefore, the Mach number should be the parameter which controls the jet strength and advance speed. This is indeed the case as shown in Fig.~\ref{p_M2_M4_M8_M12_M24_M48_rho_1}  for the density ratio of 0.1.

At Mach numbers below 4, the cocoon is stripped from the jet through Kelvin-Helmholtz instabilities. This confirms the 2D results first discussed by \citet{1982A&A...113..285N} and  the work done by \citet{2008A&A...488..795R} wherein lower Mach numbers will result in FR-I types due to the deceleration of jet material closer to the AGN source. 

At high Mach numbers, the  entire structure reaches high aspect ratios. At Mach numbers in excess of 8,  the feedback effect from the cocoon becomes increasingly evident. However, the dynamical time for the ambient medium is just  $t_o = 1 $  (3.64~Myr for the compact system). This implies the high Mach number flows penetrate all the way through the ambient medium before sound signals can cross a distance equal to the jet radius.  In this case, the cocoon has insufficient time to expand and the pressure is high. Consequently, there is high pressure feedback from the cocoon which squeezes on the jet before the terminating hot spot as these jets are clearly still in the initial blast phase.

In summary, the Mach number has a profound influence on the morphology of a radio galaxy. High Mach numbers generate highly aerodynamic and symmetric sources as well as narrow X-ray cavities. Low Mach numbers generate asymmetric turbulent lobes.

\subsection{Parameter Study: Precession}
\label{precession}

With the precession-induced jet wobbling, as defined by Equations \ref{eq_initvelocity} \& \ref{eq_velocity}, new structure occurs. The most obvious is the greater amount of mixing that occurs as compared to a relative straight jet. This is achieved through both of the defining parameters, the first being the angle at which the jet precesses and the second is the rate the jet precesses.

\begin{figure*}
\includegraphics[width=0.9\textwidth]{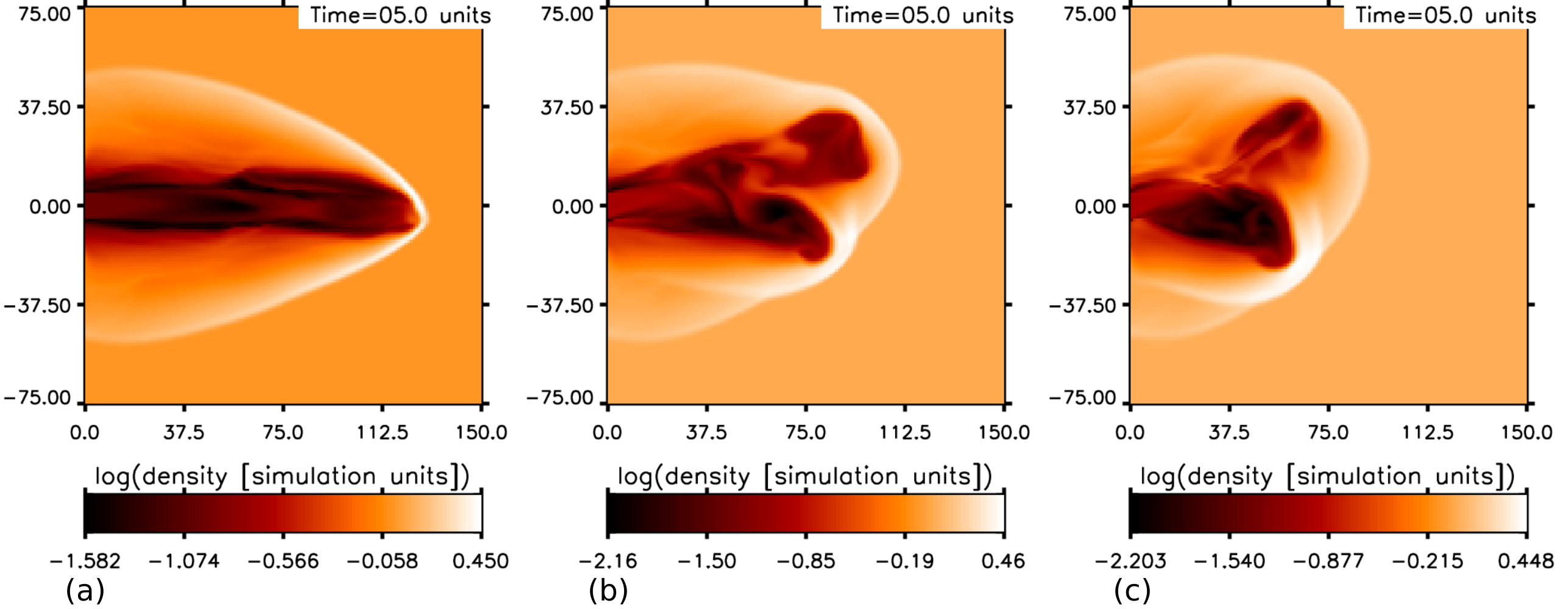}
\caption{{\small Precession dependence. Density volumetric slices from the ZEUS code with equivalent conditions bar from the precession of 1\degsy (a), 10\degsy (b) and 20\degsy (c). }}
\label{dz_1drg_10drg_20drg_rho_1}
\end{figure*}

\begin{figure*}
\includegraphics[width=0.9\textwidth]{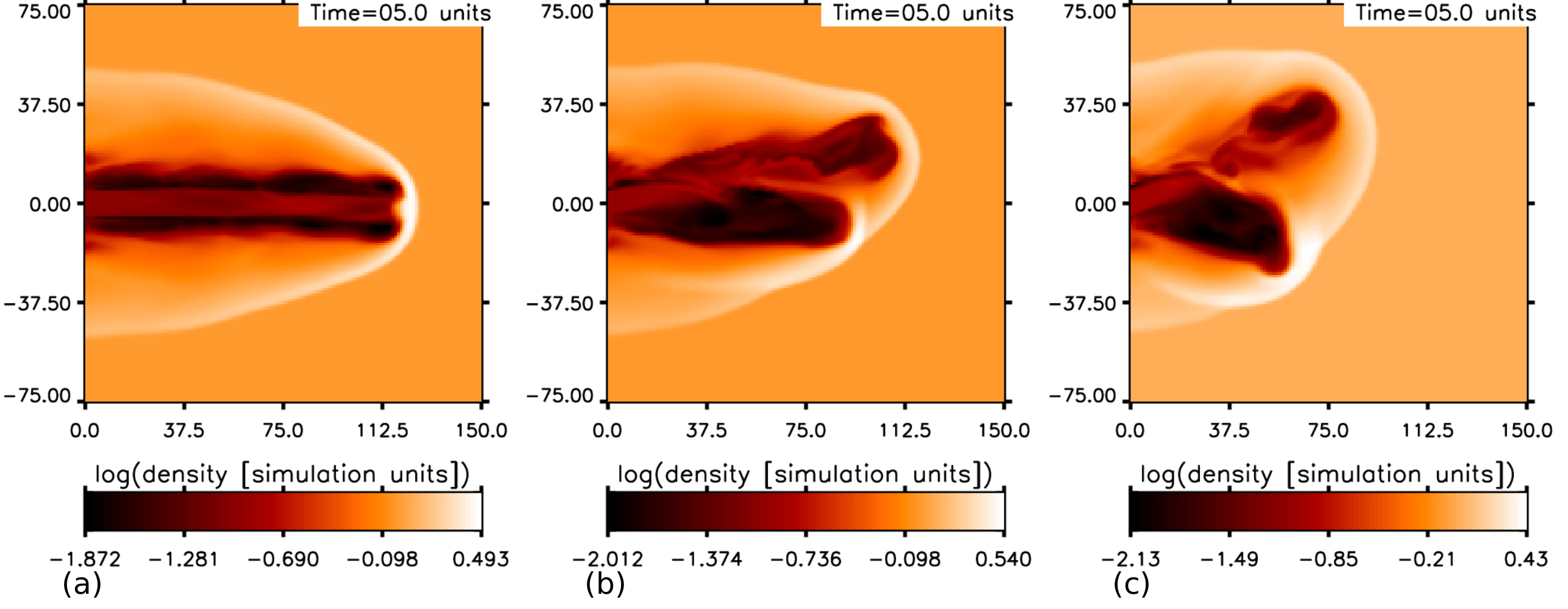}
\caption{\small{ Precession dependence. Density volumetric slices from PLUTO code with equivalent conditions bar from the precession of 1\degsy (a), 10\degsy (b) and 20\degsy (c).}}
\label{p_1drg_10drg_20drg_rho_1}
\end{figure*}

The overall structure of the lobe/cocoon changes as the angle at which the jet precesses increases, as shown in Figs.~ \ref{dz_1drg_10drg_20drg_rho_1} \& \ref{p_1drg_10drg_20drg_rho_1}.  The angle spreads the momentum and so limits the distance the jet propagates with time. It also causes the cocoon of the jet to expand and increases the total amount of mixing that can occur.
The two figures establish that the code employed has minimal influence on the structure generated once a large  precession angle is applied. The initial state is only important for  the one degree precession case where the precise numerical nozzle shape, turbulence and instability dominate over dynamical conditions.

\begin{figure*}
\includegraphics[width=0.9\textwidth]{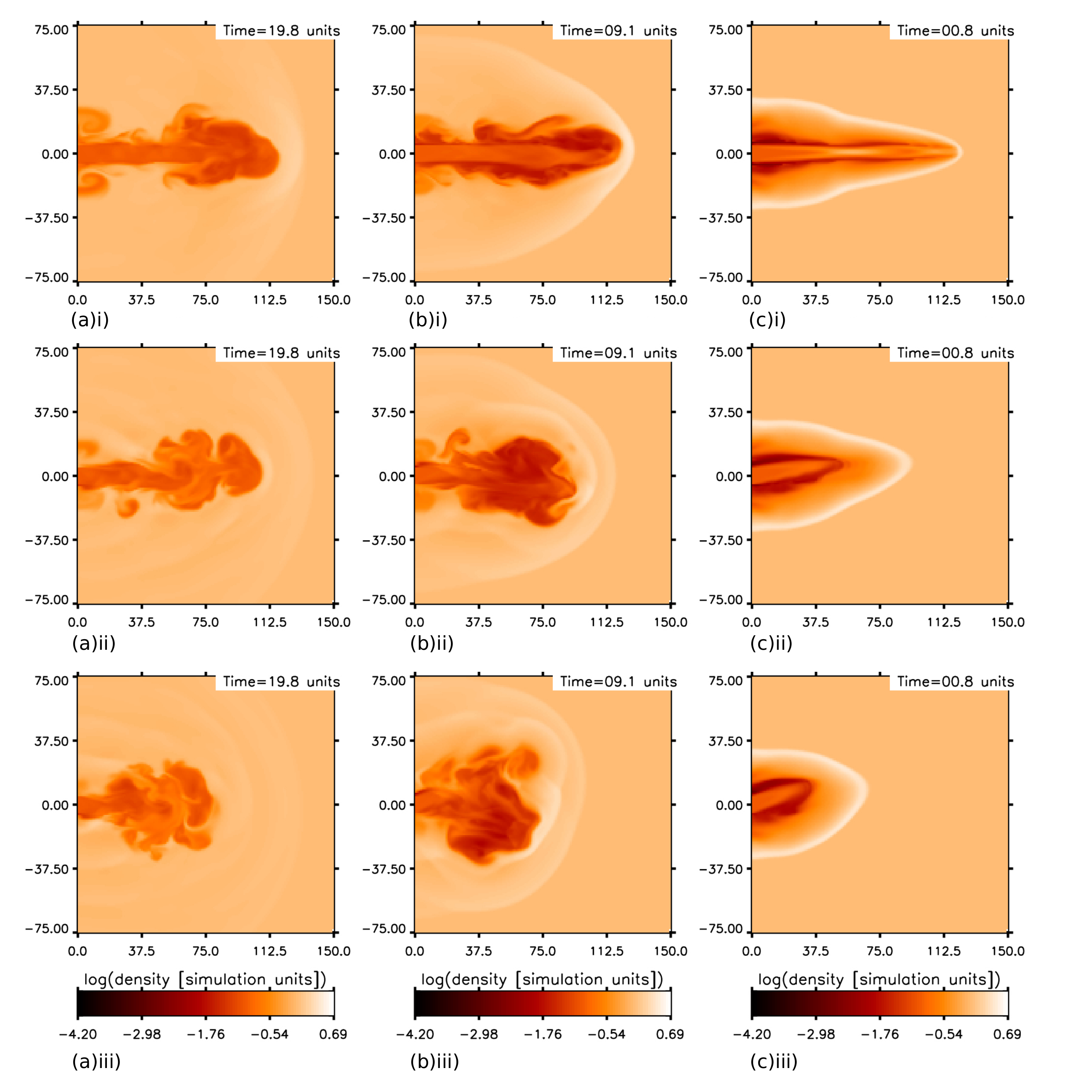}
\caption{\small{ Precession and Mach number dependence. Density volumetric slices from  the PLUTO CODE with equivalent conditions bar from the columns showing Mach numbers of 2 (a), 4 (b) and (24) and the rows showing the effect of precession of 1\degsy (i), 10\degsy (ii) and 20\degsy (iii). }}
\label{p_1drg_10drg_20drg_m2_m4_m24_rho_1}
\end{figure*}

A low rate of precession in comparison to the dynamical time generates structure similar to the standard straight jet simulation 
(e.g. right panels of Fig.~\ref{p_1drg_10drg_20drg_m2_m4_m24_rho_1}). A high precession rate relative to the dynamical time is the leading cause of the change of morphology of the jets. The standard period of precession of 4 simulation units implies that there is only just over one complete turn of the jet over the typical propagation time of the structure while the jet gas takes only $D/v_{jet}$ = 1.7 units for the jet-ambient density ratio of 0.1 and Mach number of 6. Hence, while the cocoon is distorted, the jets do not display signatures of a corkscrew or helical structure. 

As the jet precesses, new parts are coming into contact with older ejected parts of the jet as it impacts the ambient material, as seen from the multiple bow shocks in Figures \ref{dz_1drg_10drg_20drg_rho_1} \& \ref{p_1drg_10drg_20drg_rho_1}. This is causing a distorted but continuous bow shock to trace where the jet material is coming into contact with denser material. As the jet then returns to a previously excavated cavity, it has to expend less energy as the cavity is also expanding towards the ambient medium. This then results in the cocoon being made of stacked layers of jet material rather than being just the result of Kelvin-Helmholtz instabilities.

As the precession angle increases there is also an increasing cone-shaped indented region at the symmetric head of the jet, the point at which there is "symmetry" around the centre of precession. This region diminishes over time depending on the rate the jet precesses. This region plays an interesting role with precessing jets. As the jet precesses we are left with a void that traces the wake. This low pressure region is then filled in by surrounding material. This material can come from the cone region effectively detaching parts of the cocoon from the main structure. 

\begin{figure*}
\includegraphics[width=1.0\textwidth]{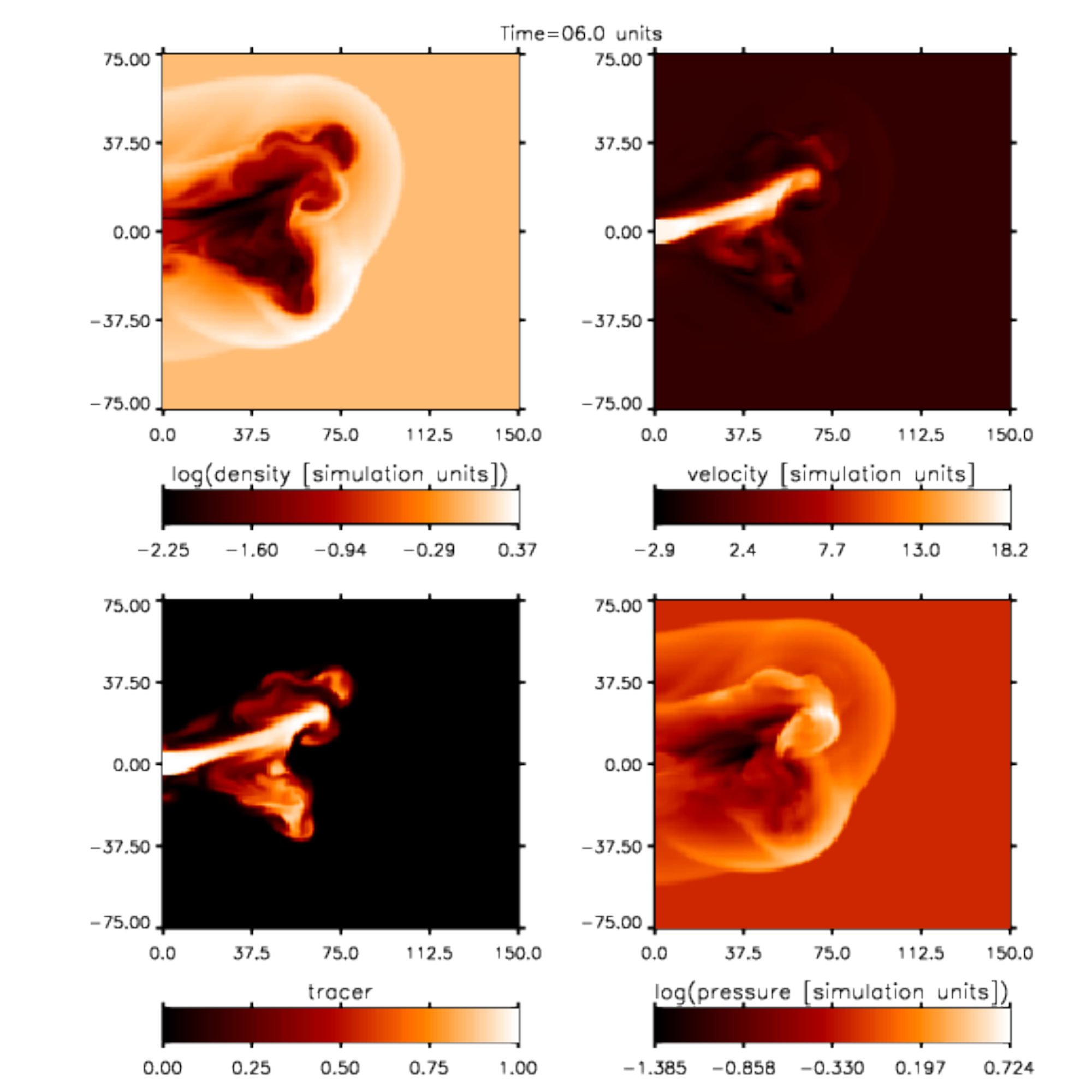}
\caption{A mid-plane slice of the parameters of a simulation with a density ratio of 0.1 with 20\degsy precession every four simulation units. Other variables are as defined in Table~\ref{parameters}.}
\label{p_jet_parameters_20xyz1_60}
\end{figure*}

Remarkably, two bow shocks can be distinguished in the ambient medium. The first is the standard advancing bow shock which envelops the entire outflow. The second following bow shock is associated with the present point of impact of the jet. This inner bow is narrower but seen prominently in the pressure panel of Fig.~\ref{p_jet_parameters_20xyz1_60} and is confirmed as pure ambient material via the tracer panel.  We thus expect such slow precession to lead to interesting radio and X-ray structures.

These effects are highlighted when studying the precession at high Mach numbers. Figure \ref{p_1drg_10drg_20drg_m2_m4_m24_rho_1} takes a sample of Mach number from Fig. (\ref{p_M2_M4_M8_M12_M24_M48_rho_1}) and shows how precession confines the head of the jet closer to the source of the material. The Mach 24 jet, column 'c' on Figure \ref{p_1drg_10drg_20drg_m2_m4_m24_rho_1}, being one of the upper end Mach numbers, has its propagation length roughly halved when precessed from 1\degsy to 20\degsy. On the other hand, the low Mach number cases displayed in column `a', display a trend of increasing break-up and plume flow as the precession angle increases.

\section{Energy}
\label{energy}

The distributions of the energy in a jet-driven system has been studied  by many authors in order to provide evidence on how radio galaxies and the ambient media evolve. One concern is the  dynamical state of the lobes which could correspond to pressure-confined reservoirs  or self-similar pressurised expansions.  A second issue is the degree of support that the jets may provide on  the entire surrounding medium through momentum driving and heating, possibly providing periodic support on a cluster-scale cooling flow. Unfortunately, there has been no systematic study to determine the general behaviour.

From two dimensional simulations,  \citet{2005A&A...429..399Z} found that up to 75\% of the injected energy is deposited into heat in the cluster gas,  although much of this would appear to be at a  late stage after the jet is switched off, while only 15\% is thermalised according to \citet{2005ApJ...633..717O} in the early phases. \citet{2013MNRAS.430..174H} demonstrated an approximate pressure balance of the lobes and ambient medium. As expected, thermal energy again dominates the energy budget although kinetic energy still accounts for  up 30\%.

\citet{2005ApJ...633..717O} summarised previous two-dimensional simulations and presented four new  three-dimensional  simulations with similar conditions  to those imposed here but with a density ratio limited to  $\rho_\text{jet}/\rho_\text{amb} = 0.01$ and a 3$^\circ$ degree conical precession. They found that approximately 40\%-60\% of the jet energy is transferred to the ambient medium with approximately 40\%-45\% of the jet energy being converted to ambient thermal energy. 

A few non-steady but axisymmetric jets were similarly studied by \citet{2010ApJ...710..180O} with the same density ratio. They reported that that  these jets typically transfer  over 60\% of their energy  into ambient thermal energy.  However, this energy is not spread wide but remains in the vicinity of the cocoon boundaries. The consequence is that such jets cannot be responsible for reheating cluster cores.

\begin{figure*}
\includegraphics[width=1.0\textwidth]{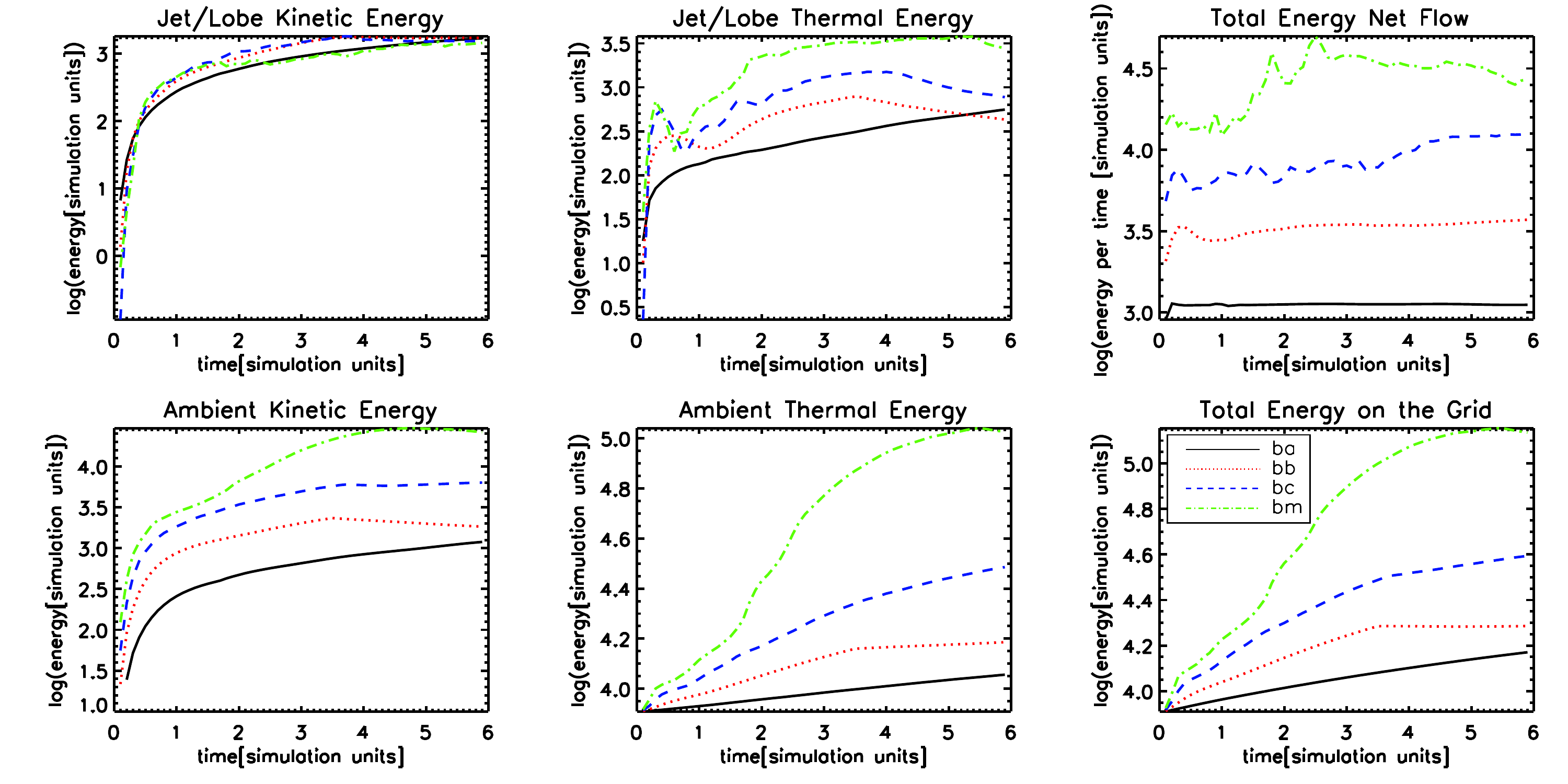}
\caption{{\small The energy distribution as it evolves with time for the four lobe/jet density ratios where the density ratios of the lines denoted ba, bb, bc and bm are 0.1, 0.01, 0.001 and 0.0001, respectively. All these simulations have outflow boundaries on all sides.}}
\label{p_ba_bb_bc_bm_energy_time}  
\end{figure*}

Figure~\ref{p_ba_bb_bc_bm_energy_time}  
shows how the bulk flow and thermal energy from the jet is redistributed as a function of the density ratio on taking a straight jet with 1$^\circ$ precession. As shown in the lower-right panel, the total net injected jet energy is not strictly proportional to the time. Instead, an initial setup phase, considerable backflow and, finally, the exit of the leading bow modify the total energy on the grid.  Even in this case we find that between 50\% and 82\% of the energy is converted into thermal energy of the ambient medium. Moreover the percentage rises as the jet-ambient density ratio falls. Thus, our results are consistent with previous findings and extends them to lower density ratios. 

Almost all the remaining energy is also transferred to the ambient medium in the form of kinetic energy. Only a few per cent remains as thermal lobe energy although there is an approximate pressure balance between the lobe and the ambient medium. 

\begin{figure*}
\includegraphics[width=1.0\textwidth]{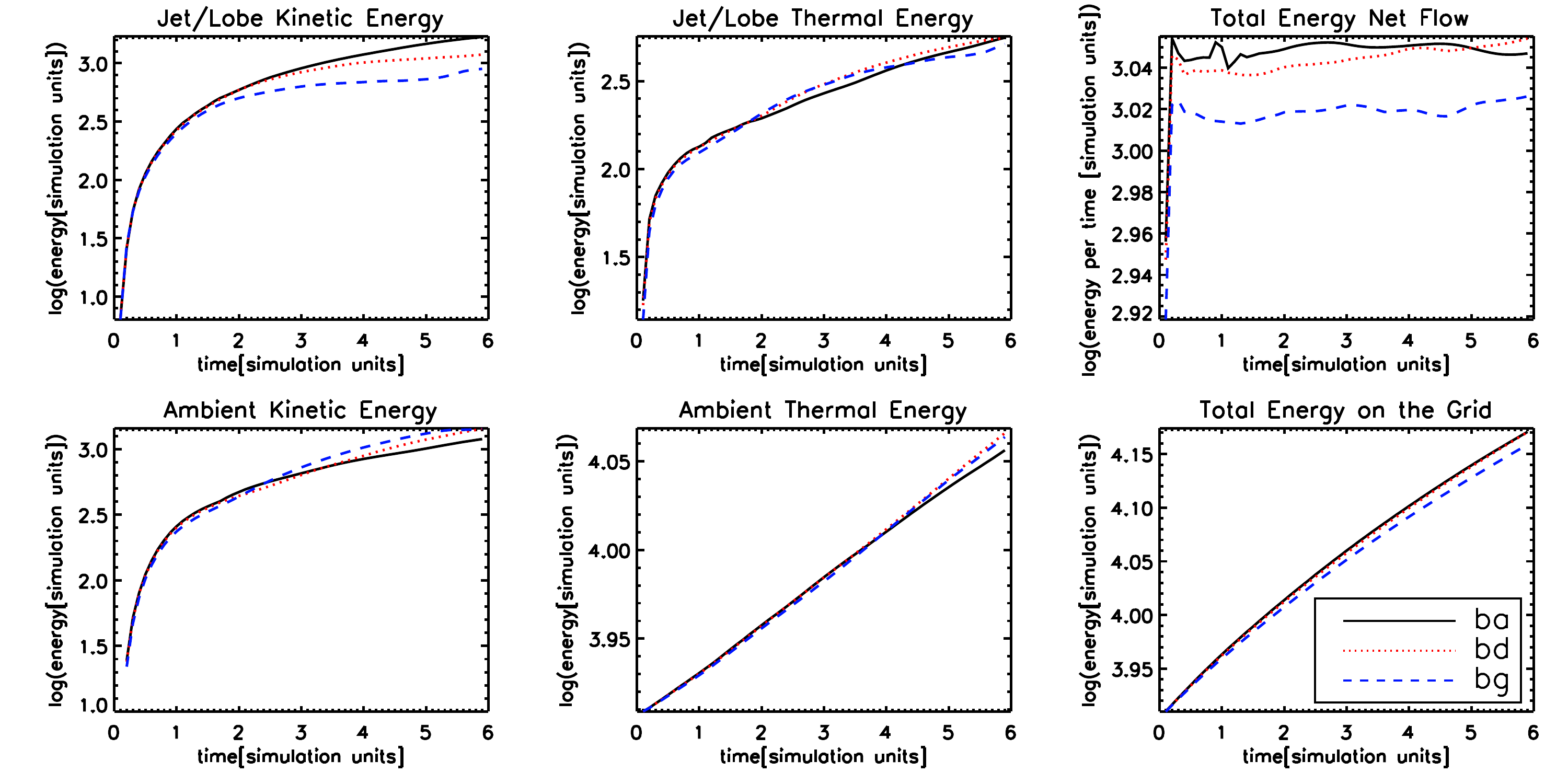}
\caption{{\small The energy distribution as it evolves with time for the the precession angles of (ba) 1\degsy, (bd)  10\degsy  and (bg) 20\degsy as denoted in the lower-right panel. 
This is the simulations with the density ratio of 0.1 and  outflow boundaries on all sides.}}
\label{p_ba_bd_bg_energy_time}  
\end{figure*}

The results carry directly over to wide precession simulations as shown in Fig.\ref{p_ba_bd_bg_energy_time}.  The main variation is in the kinetic energy of the lobe since these are small fractions which implies plenty of room for variation. With precession, somewhat more kinetic energy is contained in the ambient medium and less in the lobes which is consistent with the stunted growth of the jet itself and the spread of the impact onto the ambient medium. This time of the drop in lobe kinetic energy corresponds to the first full period of the precession. It is causing the jet/lobe thermal energy to increase as more mixing and a series of secondary shocks occurs, which is seen in figure (\ref{p_1drg_10drg_20drg_m2_m4_m24_rho_1}) for jets that complete one full precession period. 

\begin{figure*}
\includegraphics[width=1.0\textwidth]{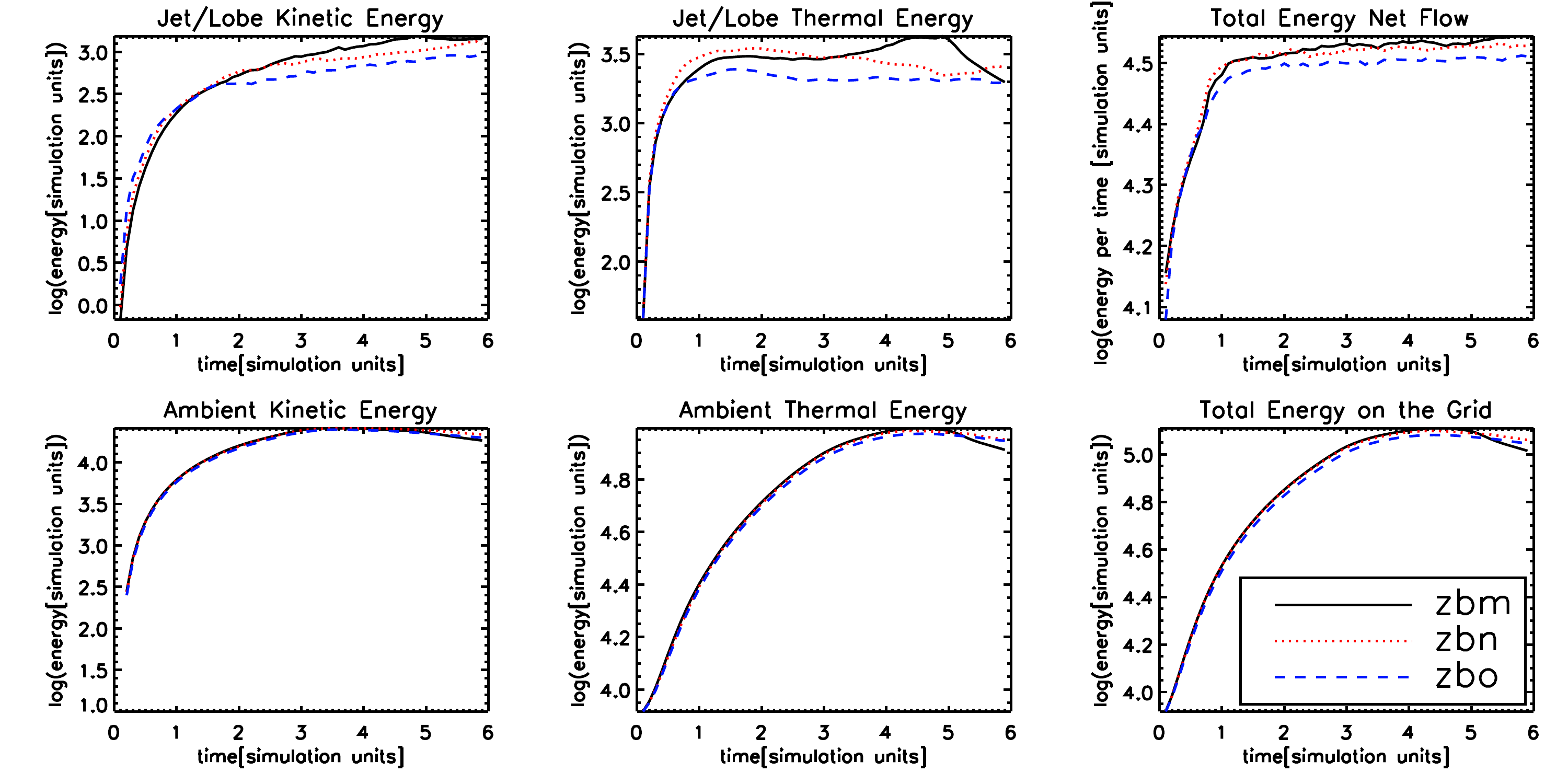}
\caption{{\small 
The energy distribution as it evolves with time for the the precession angles of (bm) 1\degsy, (bn)  10\degsy  and (bo) 20\degsy as denoted in the lower-right panel. 
This is the simulations with the density ratio of 0.0001 and  reflection boundaries on the jet inflow plane.}}
\label{p_zbm_zbn_zbo_energy_time}  
\end{figure*}

\begin{figure*}
\includegraphics[width=1.0\textwidth]{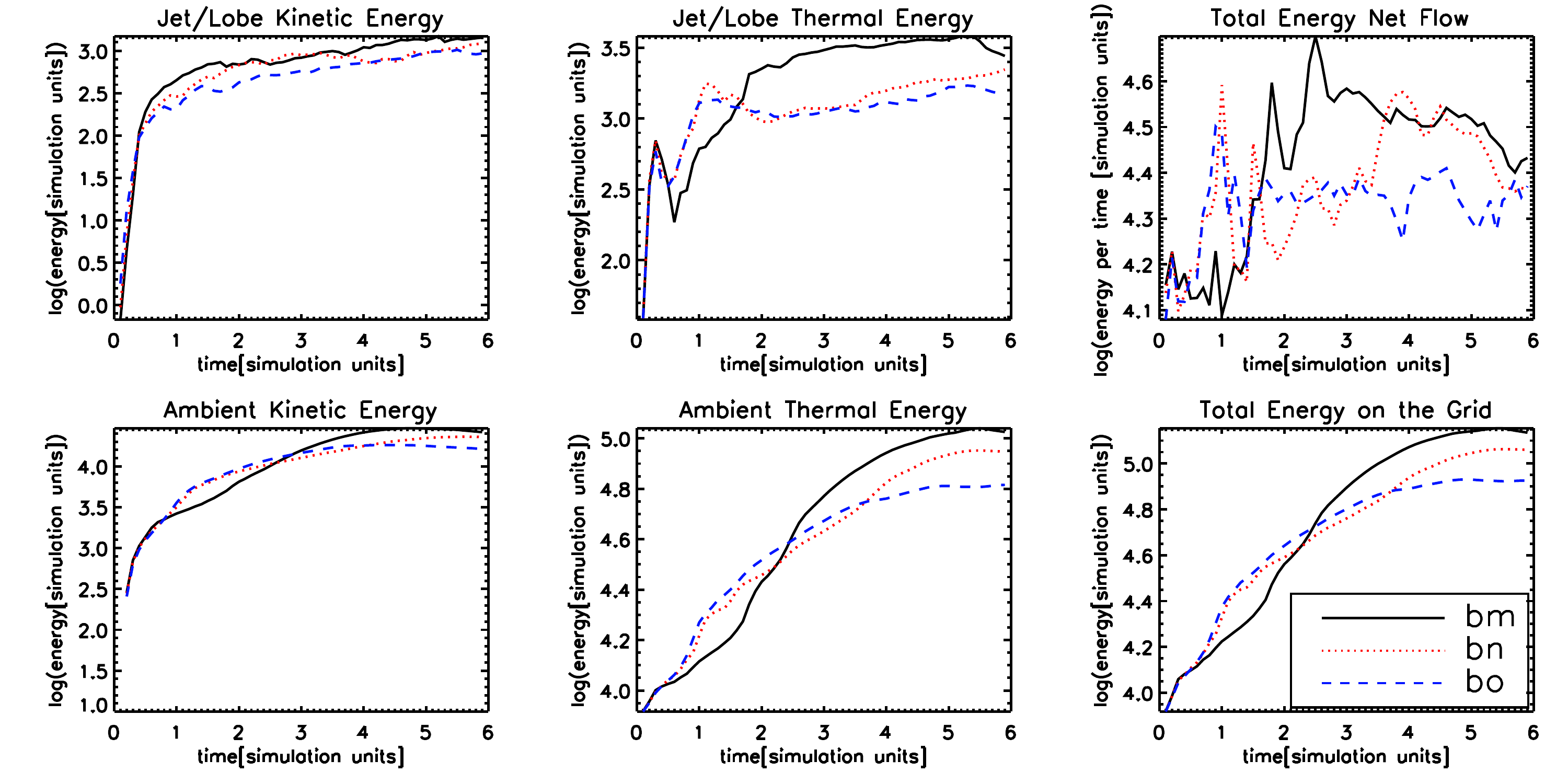}
\caption{{\small 
The energy distribution as it evolves with time for the the precession angles of (bm) 1\degsy, (bn)  10\degsy  and (bo) 20\degsy as denoted in the lower-right panel. 
This is the simulations with the density ratio of 0.0001 and an outlow boundary on the jet inflow plane.}}
\label{p_bm_bn_bo_energy_time}  
\end{figure*}

We also find that the reduction in lobe kinetic energy with increasing precession angle holds for all density ratios. However, the lobe and ambient thermal energies vary much more erratically at low jet densities, for example see Fig.~\ref{p_bm_bn_bo_energy_time}. This is due to the high sound speed within the lobes which influences both the width of the lobes and the speed of the  backflow through the lobes. We thus see that with the  reflection boundary condition on the inflow plane (Fig. \ref{p_zbm_zbn_zbo_energy_time}), the energy distributions are indeed much smoother. It can also be seen that with the outflow boundary condition the energy entering the grid increased at a slower rate during the jet setup phase but the final energy on the grid can actually be larger. This arises because, with reflection inner conditions, the ambient bow shock leaves the sides of the grid at an earlier time.


\begin{figure*}
\includegraphics[width=0.9\textwidth]{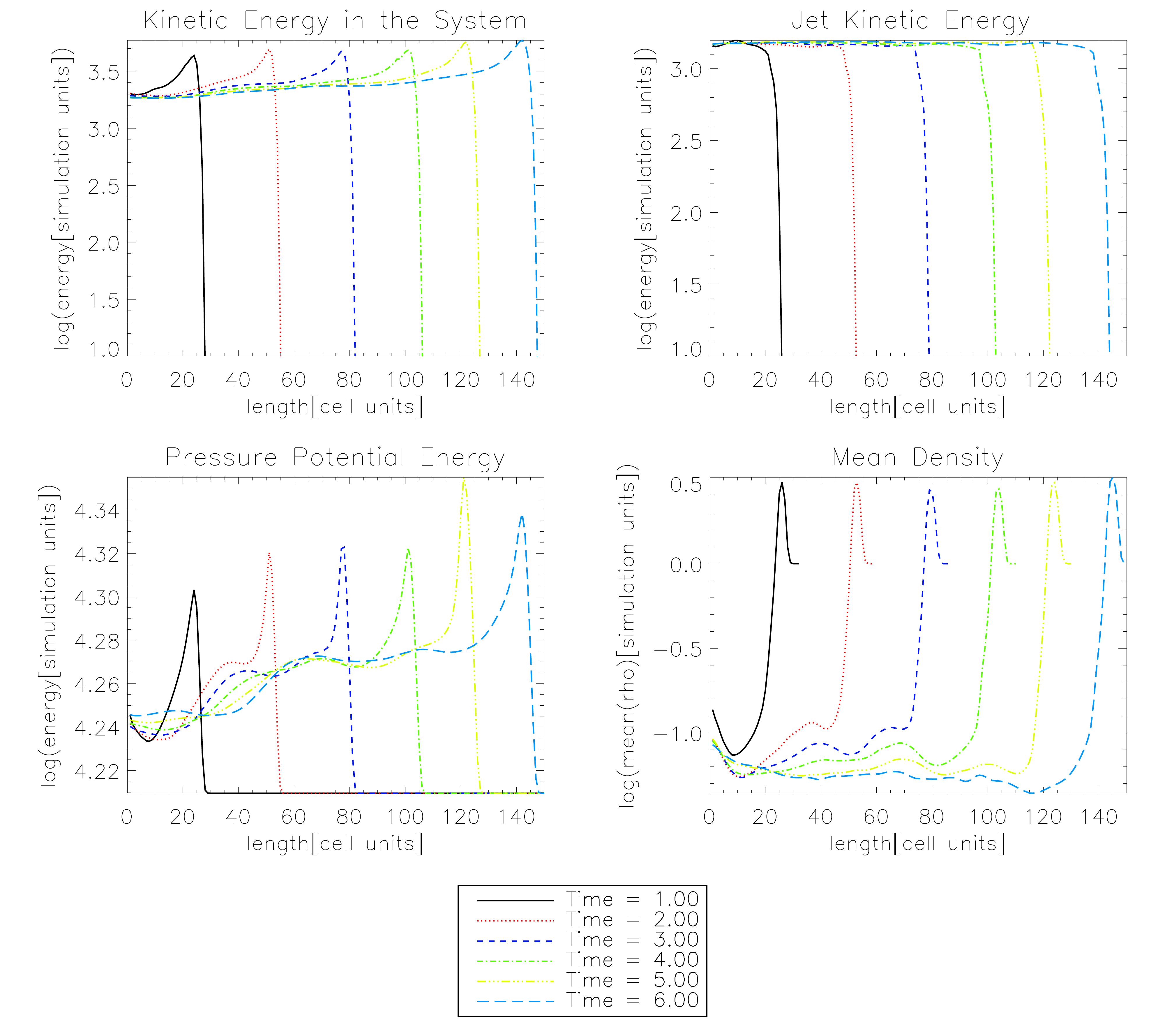}
\caption{\small{One-dimensional energy plots, integrating over  each slice perpendicular to the direction of the inflow boundary, for a Mach 6 jet with density ratio of 0.1 and precession angle of 1\degsy. Each line shows the evolution at the indicated time in simulation units.}}
\label{p_ba_10_20_30_40_50_60_energy}
\end{figure*}
\begin{figure*}
\includegraphics[width=0.9\textwidth]{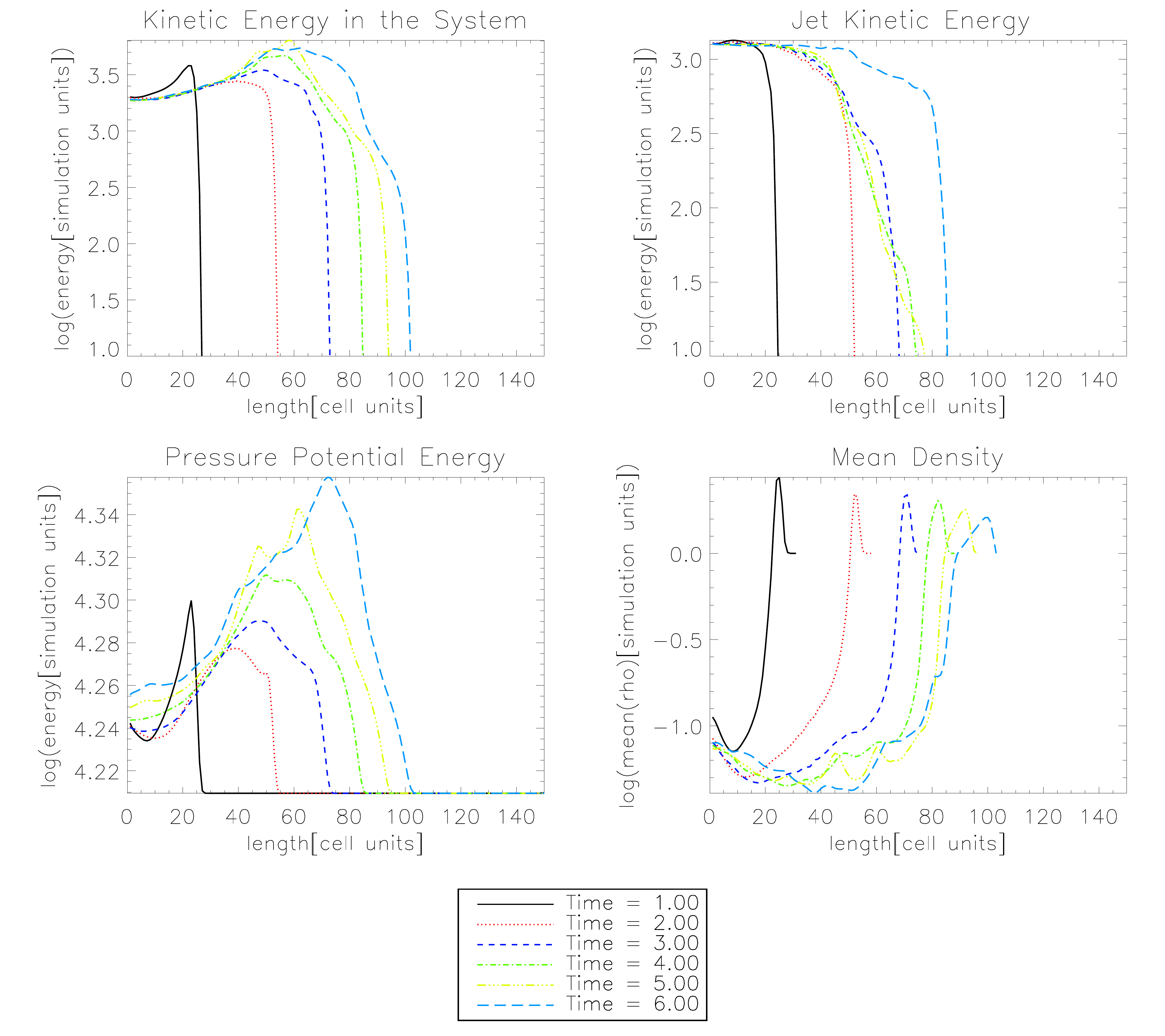}
\caption{\small{One-dimensional energy plots, integrating over each slice perpendicular to the direction of the jet propagation, for a Mach 6 jet with density ratio of 0.1 and precession angle of 20\degsy. Each line shows the evolution at the indicated time in simulation units.}}
\label{p_bg_10_20_30_40_50_60_energy}
\end{figure*}

To further analyse the effect of precession on radio galaxies, we present position-energy diagrams for the jet with a small precession (Fig.~\ref{p_ba_10_20_30_40_50_60_energy}) and with the precession angle of 20\degsy   (Fig.~\ref{p_bg_10_20_30_40_50_60_energy}). The panels display the indicated energy, pressure and density  components integrated over the slice transverse to the symmetry axis. Hence, the straight jet shows the expected one-dimensional distributions from a classical type FR{\small II} radio galaxy with a prominent leading edge. The leading edge propagates with a uniform speed. On the other hand,
the precessing examples display a gradual reduction in jet kinetic energy (top-right panel) with distance from the nozzle. The propagation speed falls systematically. However, the pressure and density are still peaked at the outer edge (lower panels) suggesting that the precession will not alter the type of radio galaxy, and we still expect these to be edge-brightened. This will be explored in detail in the following paper.

 
\section{Conclusions}
\label{conclusions}

We have performed a systematic study of supersonic hydrodynamical jets in three dimensions. These collimated jets, covering a range of low densities, are injected into a uniform ambient medium. We explore three fundamental parameters: the density, Mach number and precession angle. We present and analyse the physical structure and evolution in this work before going on to provide the implications for observed radio galaxies.

The time scale for the propagation of the radio galaxy is fixed by the Mach number, independent of the jet density for these pressure-matched jets. Nevertheless, there is a difference in the propagation speeds due to the aerodynamic shape of the impact interface as the jet density is reduced. The lowest density case, in which the jet density is 0.0001 of the ambient, does not follow this trend since the interface becomes unstable, the lobe widens and the evolution takes longer. 

As a result of the wide lobe at low densities, the shocked ambient medium is restricted to flow at high pressure close to the wings of the ambient bow shock. Therefore, the cocoon is confined by a high pressure which also acts on  the jet. The jet then behaves as an  under-pressured jet performing a convergent-divergent nozzle structure. The result is an oscillating jet within a high pressure cocoon. Therefore, distinct density conditions for which both approximate pressure-balanced and over-pressured cocoons exist. The consequences for radio emission and associated thermal X-ray emission will be explored in the following work 

The above trend with density is qualitatively reproduced in two-dimensional axisymmetric simulations (not presented here). However, the 2D jets propagate significantly slower, taking a time of
$\sim$ 8 simulation units to $\sim$  6 simulation units for the density ratios of 0.1 and 0.0001, respectively.
This is because ambient material tends to accumulate at the stagnation point which remains fixed on the flow axis in 2D but can be deflected away in 3D. 

Low Mach number jets penetrate relatively slowly through the ambient medium. This implies that there is considerable time for fluid instabilities to develop, to strip the cocoon and disturb the jet itself. At high Mach numbers, the jet initially pushes away the ambient medium which is slow to respond. The radio galaxy grows quickly and remains stable with some oscillations. The cocoon itself is highly aerodynamic and the pressure remains high in the ambient medium as the flow stays in the blast-wave phase for the first dynamical time, $t_o = 1$.

Precession is a factor requiring three dimensional simulations to study. We have shown that slow precession may generate structures with quite straight jets yet highly distorted lobe structures. As the jet precesses, new parts are coming into contact with older ejected parts of the jet as it impacts the ambient material. Low Mach number jets propagate through the ambient medium relatively slowly and so the lobe structure can be dominated by a broad impact region and wide unstable cocoons. On the other hand, we find that high Mach number jets may generate somewhat asymmetric  lobe structures. 
 
Finally, we note that a very high fraction of the grid energy is in the ambient thermal gas. We find up to 80\% for the low density jets, significantly higher than previously found for more moderate density ratios. Most of the remainder is tied up in kinetic energy of the ambient with only a few per cent available as thermal lobe energy. We thus suspect that radio galaxies driven in this manner are highly inefficient at generating the conditions of high internal energy to produce strong radio emission.
 
These simulations explore a range of basic parameters, providing knowledge which is consistent with previous studies and extends them. The PLUTO code is versatile and we can extend to MHD  relativistic flow into intergalactic and cluster environments, as has been previously studied for narrow ranges of parameters.  In particular we have found the slow precession model to be potentially of high value in interpreting radio galaxies. 

\section*{Acknowledgements}  
\label{acks}

We thank SEPNET for  part funding the Forge  supercomputer and other equipment support for JD. 

\bibliography{jets}

\label{lastpage}
\
\end{document}